\documentclass[fleqn,usenatbib,usegraphicx]{mnras}

\usepackage{newtxtext,newtxmath}

\usepackage[T1]{fontenc}

\DeclareRobustCommand{\VAN}[3]{#2}
\let\VANthebibliography\thebibliography
\def\thebibliography{\DeclareRobustCommand{\VAN}[3]{##3}\VANthebibliography}

\usepackage{graphicx}	
\usepackage{amsmath}	
\usepackage{comment}
\usepackage{float}

\title[Relic galaxy analogues in Illustris TNG50]{Relic galaxy analogues in Illustris TNG50: the formation pathways of surviving red nuggets in a cosmological simulation}

\author[Flores-Freitas et al.]{
Rodrigo Flores-Freitas,$^{1}$\thanks{E-mail: \url{rodrigoff@ufrgs.br}} 
Ana L. Chies-Santos,$^{1,2}$\thanks{E-mail: \url{ana.chies@ufrgs.br}}
Cristina Furlanetto,$^{1}$\thanks{E-mail: \url{cristina.furlanetto@ufrgs.br}}
Mar\'ia Emilia De Rossi,$^{3,4}$
\newauthor
Leonardo Ferreira,$^{5}$
Lucas J. Zenocratti,$^{6,7}$
Karla A. Alamo-Mart\'inez$^{8}$
\\
$^{1}$Instituto de F\'isica, Universidade Federal do Rio Grande do Sul, Av. Bento Gonçalves 9500, Porto Alegre, R.S. 90040-060, Brazil\\
$^{2}$Shanghai Astronomical Observatory, Chinese Academy of Sciences, 80 Nandan Road, Shanghai 200030, China\\
$^{3}$CONICET-Universidad de Buenos Aires, Instituto de Astronom\'ia y F\'isica del Espacio. Buenos Aires, Argentina\\
$^{4}$Universidad de Buenos Aires, Facultad de Ciencias Exactas y Naturales y Ciclo B\'asico Com\'un. CC 67, Suc. 28, 1428, Buenos Aires, Argentina\\
$^{5}$University of Nottingham, School of Physics \& Astronomy, Nottingham, NG7 2RD, UK\\
$^{6}$Facultad de Ciencias Astron\'omicas y Geof\'isicas, Universidad Nacional de La Plata, Paseo del Bosque s/n, B1900FWA, La Plata, Argentina\\
$^{7}$Instituto de Astrof\'isica de La Plata (IALP), UNLP - CONICET, Paseo del Bosque s/n, B1900FWA La Plata, Argentina\\
$^{8}$Departamento de Astronom\'ia, Universidad de Guanajuato, Apartado Postal 144, 36000 Guanajuato, Guanajuato, Mexico\\
}

\date{Accepted XXX. Received YYY; in original form ZZZ}

\pubyear{2021}

\begin{document}
\label{firstpage}
\pagerange{\pageref{firstpage}--\pageref{lastpage}}
\maketitle

\begin{abstract}
Relic galaxies are massive compact quiescent galaxies that formed at high-redshift and remained almost unchanged since then. In this work, we search for analogues to relic galaxies in the TNG50 cosmological simulations to understand relic formation and test the ability of TNG50 to reproduce such rare objects. Using stellar mass, age, radius, quiescence and stellar assembly criteria, we find 5 subhalos in TNG50 that could be potential relic analogues. We compare their properties with other constraints imposed by a sample of 13 observed relic galaxies. We find one analogue in TNG50 that simultaneously satisfies most of the available observational constraints, such as metallicity and morphology. It also shows similarities to the confirmed relic NGC 1277, regarding environment and dark matter fraction. By taking into account a degree of relicness, a second relic analogue may also be considered. However, the central parts of the brightness and density profiles of the analogues are less steep than that of real relic galaxies, possibly due to limited numerical resolution. We identify two formation pathways of relic analogues in TNG50 depending on their environment: they either have their remaining gas stripped during the infall into a cluster at $z \lesssim 1.2$ or consume it before $z > 1.5$. They are then deprived of significant star formation, leaving their stellar populations almost unaltered during the last 9 Gyr. We also find that the analogue progenitors at $z \sim 4$ inhabit more massive halos than progenitors of quiescent galaxies with similar stellar mass at $z \sim 0$.
\end{abstract}

\begin{keywords}
galaxies: formation -- galaxies: evolution -- galaxies: elliptical.
\end{keywords}



\section{Introduction}
Early-type galaxies (ETGs) are among the most massive galaxies known, reaching stellar masses ($M_{\rm *}$) above $10^{11}$ $\rm M_\odot$. 
Numerous studies in the past couple of decades have been dedicated to the understanding of the stellar assembly and size evolution of massive ETGs throughout cosmic time \citep{Daddi2005,Faber2007,Trujillo2007,Buitrago2008,Hopkins2009,Bezanson09,Taylor2010,vanDokkum10,Sozomoru2012,Newman2012,vanderWel14}, with many of them suggesting that ETGs observed at $z \sim 0$ have formed through the size growth of the initially compact quiescent galaxies found at higher redshifts. Some of these works indicate that such compact progenitors are in the cores of ETGs observed at $z=0$ (\citealt{Barbosa2021}). 

A currently popular formation and evolution framework for ETGs is the so-called two-phase formation scenario \citep{Bezanson09,Naab09,Oser2010,Hilz2013}. In this idealised scenario, the first phase would happen at high-redshift ($z > 2$), with the ETG progenitors undergoing wet mergers accompanied by intense star-formation \citep{Dekel2009}, growing very quickly in stellar mass - up to $M_{\rm *} \sim 10^{11} \rm M_\odot$ -  while still remaining compact in size. The outcome of this first phase would be the \textit{red nuggets}, the massive ($M_{\rm *} \sim 10^{11} \rm M_\odot$) compact quiescent galaxies observed at high-redshift \citep{Daddi2005,Damjanov2009,Schreiber2018,Valentino2020}. The second phase, in which quiescent compact galaxies would undergo dry mergers with other smaller galaxies, would explain the increase in stellar size \citep{Daddi2005,Trujillo2007,vanDokkum10}, change in internal structure and evolution of chemical gradients \citep{Spoalor2010,Cappellari2016,Martin-Navarro2018a} of massive ETGs. 
The two-phase scenario has foundations on numerical simulations of galaxy formation \citep{Bournaud2007,Naab09} and in the last decade it has been explored both observationally \citep{Taylor2010,Newman2012,Barro2013,Zibetti2020} and theoretically \citep{Oser2012,Hilz2013,Zolotov2015,wellons15,rodriguez-gomez2016}. Unfortunately, detailed direct observational studies of the first phase are still severely affected by the lack of resolution in high-redshift observations. Nonetheless, remarkable results have already been obtained for this epoch and provided insights about the formation of ETGs \citep{vanderWel2011}. 
A workaround for the current observational limitations, is to investigate analogue systems that are close to us. Given the stochastic nature of galaxy mergers, we should expect to find today in the galaxy population, a small fraction of objects that have not experienced the second phase \citep{Trujillo2009,Quilis2013,Poggianti2013b,Stringer2015}, and thus have remained unaltered since their formation, in the sense that they did not have any significant star formation episodes or mergers since $z \sim 2$. Ideally, they would remain untouched since then, allowing us to understand the properties of ETG progenitors. These analogues of the high-redshift compact quiescent galaxies, when found in the nearby Universe, are called \textit{relic galaxies} \citep{Ferre-Mateu2015}. Moreover, the study of these objects at low redshifts can complement high-redshift studies which have intrinsic spatial resolution limitations. 

In the last decade, a number of relic candidate galaxies have been identified nearby ($z \lesssim 0.05$) \citep{Trujillo2014,FerreMateu17,yildirim17,spiniello2021} and have had their relic nature verified by different works, which imposed constraints on their initial mass functions (IMF), globular cluster populations and dynamical properties \citep{Ferre-Mateu2015,martin-navarro15,Yildirim2015,yildirim17,Beasley2018,Martin-Navarro2019,Buote2019,Alamo-Martinez2021,spiniello2021,Kang2021}. 
More recently, the survey INSPIRE, is looking to confirm new relic candidates at redshifts up to $z = 0.5$ \citep{spiniello2021}.
Relic galaxy candidates should have properties similar to the {\it red nuggets} of the distant past, which means that they should be massive ($M_{\rm *} \sim 10^{11} \rm M_\odot$), compact ($R_e \leq$ 1.5 kpc), quiescent, and - since they are already evolved - very old (age $>$ 10 Gyr). 

Observationally, we can study relic galaxies in the nearby Universe to gain insight on the general processes of galaxy formation and evolution. However, we cannot directly observe their individual formation process.
Cosmological simulations offer a great opportunity to study how the properties of individual compact galaxies evolve with time \citep{wellons15,Stringer2015,wellons16,rodriguez-gomez2016,Furlong2017}, free from the constraints and biases that are present when one tries to reconstruct the history of galaxies using real observations. Nevertheless, cosmological numerical simulations also have their own limitations and biases, only recently achieving the resolution necessary to reproduce well many properties of real observed galaxies and their scaling relations \citep{Schaye2015,Dubois2016,Pillepich18b,Springel18,nelson18,Naiman2018,Genel2018,Dave2019,Vogelsberger2020}. Thus, searching for analogues of rare-type objects - such as relic galaxies - in cosmological simulations also serves as a test-bed to the numerical models themselves.

In this work, we take advantage of the IllustrisTNG simulation suite to check if one of the state-of-the-art cosmological simulations is able to generate a population of uniformly old massive compact objects, and if these objects have similar properties to the relic galaxies found in the nearby Universe. 
Inspired by the characteristics of the known relics and the selection criteria already employed in observational studies that look for massive compact quiescent galaxies \citep{Poggianti2013a,Saulder2015,Tortora2016,spiniello2021}, we search for relic analogue candidates in the IllustrisTNG simulations.
We search for old massive compact quiescent galaxies at $z=0$ in TNG50 simulations; select relic galaxy analogue candidates based on their stellar assembly; compare the properties of the selected candidates with constraints from real observations; characterise the cosmic evolution of the relic analogue candidates; determine the best relic analogue candidates and discuss their formation in the simulation. 
 As observational counterparts of relic galaxies, we use part of the sample of \cite{yildirim17} (hereafter Y17) which contains at least 13 relics galaxies. All these galaxies have already been studied in detail in other works \citep{vandenBosch2012,martin-navarro15,Yildirim2015,Beasley2018,Werner2018,Martin-Navarro2019,Buote2019,Alamo-Martinez2021,Kang2021}, allowing us to have a good characterisation of the sample, in order to constrain properties of candidate relic analogues found in TNG50. In addition to literature information, we also use Hubble Space Telescope (HST) imaging data from the Y17 sample, performing simple structural analysis.

\par This paper is structured as follows. In section \ref{sec:Data} we present the simulation data and observational sample. In section \ref{sec:Methods} we present the methods employed over the simulation data, simulation sample selection, mock images generation and photometric methods. In section \ref{sec:Results} we present the global properties, primordial halo properties, formation pathways of the best relic analogue candidates and the relics number density in TNG50.
In section \ref{sec:Conclusions} we summarise the results and present our conclusions. We adopt a solar metallicity $Z_{\rm \odot}$ = 0.0134 \citep{asplund2009} and a standard $\Lambda$CDM cosmology. The cosmological parameters are taken from \cite{Planck2016}: $H_0 = 67.74$ km~s\textsuperscript{-1}~Mpc\textsuperscript{-1}, $\Omega_m = 0.3089$ and $\Omega_\Lambda = 0.6911$, which are the same cosmological parameters adopted in the IllustrisTNG simulations.

\section{Data}\label{sec:Data}
In this section, we provide the details of the cosmological simulations and observations used throughout this work.
\subsection{Simulations}\label{sec:Simulation}
We use the publicly available suite of IllustrisTNG cosmological simulations \citep{Nelson2019a,Pillepich19}, which is a series of gravo-magnetohydrodynamical simulations evolved with the \textsc{arepo} code \citep{Weinberger2020} that incorporates a comprehensive galaxy sub-grid model. This galaxy model implement different AGN feedback modes, individual chemical element tracing and other relevant baryonic processes (check \citealt{Weinberger2017} and \citealt{Pillepich18b} for details). TNG50 is the most  computationally-demanding and highest-resolution realisation of the IllustrisTNG project, evolving dark-matter, gas, stars, black holes and magnetic fields within a uniform periodic-boundary cube of 51.7 comoving Mpc. For this work, we make use of TNG50-1 because it is the run with the highest spatial and mass resolution, enabling us to better resolve compact simulated galaxies. General characteristics of the simulation used here are presented in Table \ref{tab:simulation}.

\begin{table}
\centering
\caption{Characteristics of the TNG50-1 simulation, from top to bottom: box side-length, initial number of gas cells and dark matter particles, the target baryon mass, roughly equal to the average initial stellar particle mass, the dark matter particle mass, the $z=0$ Plummer equivalent gravitational softening of the collisionless components, the minimum comoving value of the adaptive gas gravitational softening and total run time. Data from \protect\cite{Pillepich2019}.}
\begin{tabular}{lc} \hline
Simulation                    & TNG50-1               \\
L$_{\rm box}$ [cMpc]             & 51.7                  \\
N$_{\rm gas}$                    & 2160$^3$              \\
N$_{\rm DM}$                     & 2160$^3$              \\
m$_{\rm baryon}$ [$\rm M_\odot$] & 8.5$\times 10^4$      \\
m$_{\rm DM}$ [$\rm M_\odot$]     & 4.5$\times 10^5$      \\
$\varepsilon_{\rm *}$($z=0$) [pc]  & 288                   \\
$\varepsilon_{\rm gas,min}$ [pc] & 72                    \\
CPU Time [h]                 & $\sim 130 \times 10^6$ \\ \hline
\end{tabular}
\label{tab:simulation}
\end{table}

The information about the particles inside the simulation is stored in hierarchical files called "snapshots", each corresponding to a specific time in the virtual universe. These snapshots carry positions, velocities and other properties for each particle in the entire TNG50 box. The larger structures inside the simulation are identified using a friends-of-friends algorithm (FoF), which recognises what are called "FoF halos" - or simply halos - based on over-densities of dark matter particles. Subsequently, the \textsc{subfind} algorithm \citep{Springel2001,Dolag2009} detects substructures in these halos which are them referred to as "subhalos": the structures of gravitationally bound particles in the simulation that are analogous to galaxies. The properties of each halo and subhalo are available in group catalogues generated with these algorithms. 
\par In this work, we make use of snapshot data, group catalogues and supplementary data catalogues\footnote{\url{https://www.tng-project.org/data/}}, the latter being generated by post-processing of snapshot data. To access the data and partly process it, we use the python library \textsc{illustris\_python} and a JupyterLab work-space, hosted and provided by the TNG collaboration. 
In order to analyse in more detail the gas dynamics of the simulated galaxies, we also downloaded and processed snapshot data from one of the subboxes of TNG50-1\footnote{Videos produced with this data are available online in the Supplementary Material of this publication.}. The subboxes in TNG simulations are fixed regions in the larger simulation box which have snapshot data with higher time resolution, allowing us to analyse in more detail the internal dynamics of some subhalos. To identify and trace the subhalos in the subboxes, we use the supplementary catalogue created by \cite{Nelson2019a}. We also use the stellar assembly supplementary catalogue of \cite{rodriguez-gomez2016} to determine stellar mass \textit{in-situ} fractions.

\subsection{Observations}\label{sec:obs_sample}
In this section, we briefly describe the observational sample used as a reference to impose constraints on the properties of relic analogue candidates found in the TNG50 simulation.
The observational sample is composed of 13 nearby massive compact elliptical galaxies (CEGs) which are presented in \cite{yildirim17} and further considered as relic galaxies in other works. The original Y17 sample is considerably biased towards objects with high-velocity dispersion and is composed of only 16 galaxies. In order to avoid ambiguity regarding the relic nature of galaxies in our observational sample, due to signs of relatively younger stellar populations or recent interaction, 3 galaxies from Y17 sample are not considered in this work (NGC 1282, NGC 3990 and UGC 3816). For the photometric analysis, we use drizzled H-band images (HST F160W filter) retrieved from the Hubble Legacy Archive (HLA), programme GO: 13050 (PI: van den Bosch). In order to have a homogeneous photometric analysis, we only consider the images of 11 galaxies (see Fig. \ref{fig:H-band}). NGC 1277 simply lacks HST H-band imaging, while PGC 12562 drizzled images have failed guide star and do not have the same exposure time or pixel scale as the rest of the sample, therefore we do not present photometric analysis for them here. In Table \ref{tab:obs_properties}, we present some relevant properties of the observational sample.

\begin{table*}
\centering
\caption{Properties of the observational sample, from left to right: galaxy identifier, stellar mass, dark matter-to-stellar mass ratio, elliptical half-light radius, stellar metallicity, age, central velocity dispersion - all from \protect\cite{yildirim17} - and adopted distance.}
\begin{tabular}{ccccccccccc} \hline
ID        & log$(M_{\rm *}/\rm M_\odot)$ & log$(M_{\rm DM}/M_{\rm *})$ & $R_{\rm e,ell}$ & log$(Z_{\rm *}/Z_{\rm \odot})$ & Age & $\sigma_{\rm c}$ & D \\
          &       & & [kpc] &       & [Gyr] & [km s$^{-1}$] & [Mpc] \\ \hline
MRK 1216  & 11.34$^{+0.11}_{-0.10}$ & 2.47$^{+2.68}_{-5.11}$ & 3.0 $\pm$ 0.1 & - & - & 335 $\pm$ 6 & 96 $\pm$ 2\\
NGC 0384  & 10.96$^{+0.05}_{-0.05}$ & 1.86$^{+0.84}_{-0.68}$ & 1.8 $\pm$ 0.1 & 0.05 $\pm$ 0.05 & 13.7 $\pm$ 0.36 & 240 $\pm$ 5 & 62 $\pm$ 2\\
NGC 0472  & 11.07$^{+0.06}_{-0.11}$ & 1.36$^{+1.35}_{-0.51}$ & 2.4 $\pm$ 0.1 & 0.03 $\pm$ 0.03 & 13.4 $\pm$ 0.31 & 252 $\pm$ 7 & 79 $\pm$ 2\\
NGC 1270  & 11.31$^{+0.10}_{-0.12}$ & 1.93$^{+1.49}_{-3.96}$ & 2.2 $\pm$ 0.1 & 0.34 $\pm$ 0.02 & 14.0 $\pm$ 0.50 & 376 $\pm$ 9 & 74 $\pm$ 2\\
NGC 1271  & 11.06$^{+0.07}_{-0.07}$ & 1.92$^{+0.65}_{-0.66}$ & 2.0 $\pm$ 0.1 & 0.13 $\pm$ 0.02 & 14.0 $\pm$ 0.50 & 302 $\pm$ 8 & 89 $\pm$ 2\\
NGC 1277  & 11.13$^{+0.06}_{-0.07}$ & -0.91$^{+3.58}_{-1.44}$ & 1.3 $\pm$ 0.1 & - & - & 355 $\pm$ 5 & 75 $\pm$ 2\\
NGC 1281  & 11.00$^{+0.08}_{-0.08}$ & 1.49$^{+1.57}_{-2.54}$ & 1.6 $\pm$ 0.1 & 0.21 $\pm$ 0.03 & 14.0 $\pm$ 0.50 & 263 $\pm$ 6 & 63 $\pm$ 2 \\
NGC 2767  & 11.12$^{+0.09}_{-0.08}$ & 1.54$^{+0.83}_{-3.52}$ & 2.4 $\pm$ 0.1 & 0.11 $\pm$ 0.02 & 14.0 $\pm$ 0.50 & 247 $\pm$ 9 & 74 $\pm$ 2\\
PGC 11179 & 11.16$^{+0.06}_{-0.08}$ & 0.81$^{+1.97}_{-3.00}$ & 2.1 $\pm$ 0.1 & 0.11 $\pm$ 0.03 & 14.0 $\pm$ 0.50 & 292 $\pm$ 7 & 103 $\pm$ 2\\
PGC 12562 & 10.74$^{+0.10}_{-0.09}$ & 2.92$^{+1.08}_{-4.93}$ & 0.7 $\pm$ 0.1 & 0.23 $\pm$ 0.03 & 14.0 $\pm$ 0.50 & 260 $\pm$ 7 & 70 $\pm$ 2\\
PGC 32873 & 11.28$^{+0.04}_{-0.04}$ & 2.47$^{+0.48}_{-0.46}$ & 2.3 $\pm$ 0.1 & 0.27 $\pm$ 0.02 & 14.0 $\pm$ 0.50 & 308 $\pm$ 9 & 113 $\pm$ 2\\
PGC 70520 & 10.95$^{+0.10}_{-0.12}$ & 2.85$^{+0.87}_{-0.75}$ & 1.6 $\pm$ 0.1 & 0.09 $\pm$ 0.02 & 14.0 $\pm$ 0.50 & 259 $\pm$ 8 & 75 $\pm$ 2\\
UGC 2698  & 11.58$^{+0.01}_{-0.03}$ & -1.05$^{+1.32}_{-1.32}$ & 3.7 $\pm$ 0.1 & 0.20 $\pm$ 0.02 & 14.0 $\pm$ 0.50 & 351 $\pm$ 8 & 95 $\pm$ 2\\ \hline
\end{tabular}
\label{tab:obs_properties}
\end{table*}

The distances adopted to these galaxies are luminosity distances computed from their redshifts. As well as the physical scale (kpc/arcsec), the distances are computed with the cosmology module from \textsc{astropy}, using the \texttt{FlatLambdaCDM} object with the cosmological parameters adopted in this work. The redshift for each galaxy was taken from the NASA Extragalactic Database, being the redshift measurement with the smallest uncertainty.

\section{Methods}\label{sec:Methods}
In this section, we outline the methods employed to select and analyse relic analogues in TNG50.
\subsection{Simulation methods and definitions}
In this subsection, we describe the fiducial choices adopted and the methods applied for the analysis of the simulation data. Stellar mass ($M_{\rm *}$), star-formation rate (SFR) and other relevant quantities for this work are measured in two different 3D spherical apertures: inside 2 stellar half-mass radius\footnote{This quantity is computed in 3 dimensions, it corresponds to the radius of a sphere containing half of the total stellar mass in a subhalo.} ($R_{\rm e,*}$) and inside 30 kpc. The first aperture is commonly used in TNG-related works and the latter is commonly used in studies regarding the EAGLE simulations, being associated with the 2D Petrosian apertures that are frequently used in observational studies \citep{Schaye2015}. We adopt the centre of the subhalos as the position of the most bound particle\footnote{The particle with the minimum gravitational potential energy.} and throughout the paper, we adopt the 30 kpc aperture as the fiducial choice, using the 2$R_{\rm e,*}$ aperture for comparisons and sample selection tests.
\par The group catalogue of TNG50 provides instantaneous SFR derived from the gas cells in each subhalo, but in order to use a quantity closer to real observables in galaxies, during the sample selection we compute the SFR in finite timescales, similar to what is done in previous works using TNG data \citep{Donnari2019}. We simply sum the initial masses of the stellar particles formed in the previous 200 Myr and 1000 Myr - relative to the snapshot of interest - and divide it by the respective timescale. For the purpose of this work, we adopt the timescale of 1000 Myr, which we consider appropriate for the selection of long quiescent objects at $z \sim 0$. Quiescent objects are selected as subhalos with sSFR$\leq 10^{-11}$ yr\textsuperscript{-1}.
\par The mass-weighted age and metallicity profiles are constructed using spherical shells with radial bins 0.2 kpc wide, evenly spaced, from $R_{\rm min}$ = 0.05 kpc to $R_{\rm max}$ = 40 kpc, computing median values for the quantities in bins which have at least 100 stellar particles. We use the same radial bins and limits for the stellar surface density profiles, which depend on the orientation with respect to the line-of-sight. Since all the galaxies analysed here have unambiguous rotation disks (see Fig. \ref{fig:kinematics} in section \ref{results:kinematics}), for simplicity, we reorient the objects to be face-on, then we compute the surface densities in circular annuli. We perform rotations in the reference frame of the subhalos so that their principal axes become parallel to the cartesian axes of a new reference frame, with the intrinsic minor axis of the subhalo being parallel to the z-axis. The rotation matrices are obtained by diagonalization of the moment of inertia tensor, computed using the stellar particles within 2$R_{\rm e,*}$, following a similar method already employed in other works which use TNG data \citep{Pillepich19,Pulsoni2020}. We also compute stellar surface densities inside 1$R_{\rm e,*}$ ($\Sigma_{\rm e,*}$) as a compactness metric. We simply use the following equation: 
\begin{equation}
    \Sigma_{\rm e,*} = \frac{M_{\rm *}(<R_{\rm e,*})}{\pi R_{\rm e,*}^2},
\end{equation}
where $M_{\rm *}(<R_{\rm e,*})$ is the stellar mass inside an spherical region of radius $R=R_{\rm e,*}$. In order to check the consistency with projected observational measures, we also estimated similar quantities considering subhalo particles within cylinders with radial aperture $R_{\rm e,*}$ and different random axial orientations (see Supplementary Material). Our main trends and findings are preserved when calculating properties inside such cylinders, instead of in a spherical region.
\par We also perform a stellar kinematics analysis, by building kinematic maps using square pixels. We do not generate mock integral field unit observations, instead we compute the values in each pixel directly from the line-of-sight velocities of the stellar particles, using a similar approach as other works based on TNG data \citep{Pillepich19,Pulsoni2020}. As a fiducial choice, we use square pixels with 0.5 kpc side, which is roughly comparable to the resolution of the kinematic maps presented in Y17. Maps with higher resolution have also been computed and are available as supplementary material. For each $i$-th pixel in our maps, we compute weighted velocities ($V_i$) and velocity dispersions ($\sigma_i$):  
\begin{equation}
    V_i = \frac{\sum_{n}^{N_i}w_n V_n}{\sum_{n}^{N_i}w_n}; \ \ \ \ \ \ \sigma_i = \frac{\sum_{n}^{N_i}w_n (V_n-V_i)^2}{\frac{N_i}{N_i-1}\sum_{n}^{N_i}w_n},
\end{equation}
where $n$ is the index for each particle inside the projection of a given pixel, $V_n$ is the line-of-sight velocity of the $n$-th particle relative to the subhalo, which has a peculiar velocity $V_{\rm sys}$, and $w_n$ is the weight used. In this work, $V_{\rm sys}$ is simply the mass-weighted sum of line-of-sight velocities of all particles in the subhalo and $w_n$ is the mass of the particles. We also compute the maps using luminosity weights, but the results do not change significantly. Moreover, we only compute the kinematic quantities in pixels which have at least $N_i=100$ particles.
\par For the cosmic evolution analysis, we use the merger trees of the simulation constructed with the \textsc{sublink} algorithm, which links progenitor subhalos to their unique descendants in future snapshots based on the particles that they have in common and a merit function that takes into account the binding energy of the particles of the descendants \citep{Rodriguez-Gomez2015}. We track the evolution of the subhalos along their main progenitor branches, computing quantities directly from the snapshot data when necessary and only analysing their properties in the time interval in which they are well resolved. We consider that a subhalo is well resolved once it has at least 5000 stellar particles and this condition is satisfied over the last 12.75 Gyr for all the objects in the final sample studied here. From the merger trees of TNG we can also obtain the number of mergers that a subhalo has already gone through. We classify the mergers by the stellar mass ratio of the galaxies involved ($M_{\rm 1}/M_{\rm 2}$). We define: major mergers if $M_{\rm 1}/M_{\rm 2} \geq 1/4$, minor mergers if $1/4 > M_{\rm 1}/M_{\rm 2} \geq 1/10$ and micro mergers if $1/10 > M_{\rm 1}/M_{\rm 2} \geq 1/100$.
\par When analysing the environment of subhalos in the simulation, we refer to the FoF halo in which a galaxy inhabit as its host halo. As the halo mass, we use its virial mass ($M_{\rm 200,c}$), which is the total mass of a sphere around the centre of the halo with a mean density of 200 times the critical density of the universe. When we declare that a simulated galaxy is a satellite of a halo, we only mean that it is assigned to this halo by the \textsc{subfind} algorithm and it is not the most massive subhalo in the halo.

\subsection{Sample selection in TNG50}\label{sec:sample_selection}
We first select all the subhalos at $z = 0$ which have $M_{\rm *} \geq 10^{9.75} \rm M_\odot$ and \texttt{SubhaloFlag}=1\footnote{This flag indicates if a subhalo is from cosmological origin (1) or not (0), in the sense that they may not have formed due to the process of structure formation and collapse. Some subhalos will represent fragments or clumps produced from already formed galaxies.}. Then, we perform multiple searches for relic analogue candidates, each with slightly different criteria, in order to contemplate different selections that could be adopted in order to select old compact massive quiescent galaxies in a hypothetical survey. For the queries applied, we consider stellar half-mass radius ($R_{\rm e,*}$), stellar mass ($M_{\rm *}$), star formation rate (SFR), specific star formation rate (sSFR) and mass-weighted age profiles, with $M_{\rm *}$, SFR and sSFR being measured in the 2$R_{\rm e,*}$ and 30 pkpc (proper kpc) apertures. 
\par For each selection quantity we adopt a threshold. For $M_{\rm *}$, we consider $ 10^{10.5} \rm M_\odot$ as the minimum value, since the observed relic galaxies are expected to have $\rm M_{\rm *} \sim 10^{11} \rm M_\odot$. By applying a threshold 0.5 dex smaller, we account for plausible existing intrinsic $M_{\rm *}$ underestimations in the simulation and also allow the possibility to find less massive relic analogues. Note that, given the limited size of the simulation box, the expected number of massive quiescent galaxies could be relatively small.
This low $M_{\rm *}$ threshold also partially takes into account subtle variations in the stellar mass value assigned to the subhalos due to the way their structures are identified by the \textsc{subfind} algorithm. 
\par For the $R_{\rm e,*}$, we first assume that it can be at maximum equal to the elliptical half-light radius ($R_{\rm e,ell}$) of real relic galaxies, a commonly used size measurement in observational studies. In the observational sample used here (from Y17), the maximum half-light radius in the H-band is $R_{\rm e,ell}$ = 3.7 $\pm$ 0.1 kpc. Other studies often classify massive galaxies as compact if their half-light radii are smaller than 2 kpc \citep{Tortora2016,buitrago2018,spiniello2021}. Here, we assume that relic analogues in the TNG50 simulation could have slightly larger sizes, so the maximum $R_{\rm e,*}$ used in the searches is 4~kpc. Secondly, we search for relic analogues without using fixed thresholds for $R_{\rm e,*}$ but selecting the most compact simulated galaxies at a given $M_{\rm *}$. More specifically, we selected subhalos below the 25th $R_{\rm e,*}$ percentile in log$(M_{\rm *})$ bins of 0.25-0.5 dex, along the interval $10^{9.75}-10^{12.5} \ \rm M_\odot$. The first (second) size criteria is depicted as $R_{\rm e,*}$ ($M_{\rm *}-R_{\rm e,*}$) in Table \ref{tab:filters}.  
\par In the case of SFR, we consider an upper limit of SFR~=~1~$\rm M_\odot$yr$^{-1}$, which, for the timescale adopted, would correspond to an increase of less than 4 per cent in stellar mass for a subhalo with the minimum mass adopted ($M_{\rm *} = 10^{10.5} \rm M_\odot$). Such value of SFR is more than 3 orders of magnitude above the smallest non-vanishing SFR value (SFR$_{\rm min}$) in TNG100-1 for the same timescale \citep{Donnari2019}. Since TNG100-1 has lower resolution than TNG50-1 and that higher mass resolution implies even smaller SFR$_{\rm min}$, the adopted SFR threshold is not significantly affected by resolution effects in the context of this work. Alternatively, for selecting quiescent galaxies we consider a maximum sSFR = $ 10^{-10.5}$yr$^{-1}$, even more flexible than the commonly adopted division value in literature - $10^{-11}$ yr$^{-1}$ - which roughly corresponds to the minimum in the bimodal distribution of sSFR, independent of the stellar mass of the galaxy \citep{Wetzel2012}.
\par Finally, we select only the subhalos that are older than 9.5 Gyr over all radial bins used to construct their profiles, thus accepting that some residual star formation could still be happening until $z \sim$ 1.5. The age profile criteria is the most restrictive one, drastically reducing the number of candidates in our queries\footnote{We repeat the searches using r-band luminosity-weighted age profiles and the final selected sample remains unchanged.}. In Table \ref{tab:filters}, we briefly list the different criteria used to search for relic galaxy analogues as described in this subsection. 
\begin{table}
\centering
\caption{Queries used to search for relic analogue candidates in TNG50, each query uses a different combination of apertures, criteria for size and quiescence selections. Additionally, the candidates are selected to be older than 9.5 Gyrs along their age profiles. The rightmost column indicates the number of subhalos selected by a given query. See the details of each criteria in section \ref{sec:sample_selection}.}
\begin{tabular}{ccccc}
\hline
Query & Aperture & Size & Quiescence & N \\ \hline
A1     & $2R_{\rm e,*}$ & $R_{\rm e,*}$ & SFR &  31\\ 
A2     & $2R_{\rm e,*}$ & $R_{\rm e,*}$ & sSFR & 31\\ 
A3     & $2R_{\rm e,*}$ & $M_{\rm *}-R_{\rm e,*}$ relation & SFR & 37\\ 
A4     & $2R_{\rm e,*}$ & $M_{\rm *}-R_{\rm e,*}$ relation & sSFR & 38\\ 
B1     & 30 pkpc & $R_{\rm e,*}$ & SFR & 23\\ 
B2     & 30 pkpc & $R_{\rm e,*}$ & sSFR & 24\\ 
B3     & 30 pkpc & $M_{\rm *}-R_{\rm e,*}$ relation & SFR & 23\\ 
B4     & 30 pkpc & $M_{\rm *}-R_{\rm e,*}$ relation & sSFR & 24\\ \hline
\end{tabular}
\label{tab:filters}
\end{table}

After applying all the queries listed in Table \ref{tab:filters}, we find a total of 48 unique relic analogue candidates. 
We decided to study simulated galaxies with stellar assembly closer to the theoretical definition of ideal relics, in order to see if their resulting properties at z = 0 match the properties of relic galaxies in observations. Thus, we proceed the selection of relic analogues by applying an additional cut over the stellar populations ages of the 48 candidates, selecting only those subhalos with more than 75 per cent of stellar mass composed by populations older than 9.5 Gyr, reducing the sample to only 11 relic analogue candidates. Then, taking advantage of the simulations, we analyse the stellar-size evolution of these 11 objects and notice that 4 of them have passed through significant and abrupt changes in their half-mass radii ($\Delta R_{\rm e,*} \gtrsim 1$~kpc) since $z \sim 1.5$, possibly due to strong interaction with their respective massive neighbours. These latter objects are thus not considered in the final sample since they could not represent relic galaxy analogues at $z=0$. We further analyse the stellar surface density and surface brightness distribution of the remaining candidates, noticing that 2 subhalos have stellar streams around them, indicating recent interactions or ongoing mergers. These 2 objects are not considered in the final analogue candidates sample. Therefore, the final sample of relic analogue candidates analysed in detail in this work is composed of 5 subhalos, with general properties presented in Table \ref{tab:properties}. We leave for a future work, the detailed study of the 48 simulated massive old compact quiescent galaxies found initially, because such objects may be part of a broader category with more diverse formation histories than relic galaxies, and thus need to be compared with a more complete observational sample.

\begin{table*}
\centering
\caption{Properties of the TNG50 final sample of relic analogue candidates at $z = 0$. For columns with 2 numerical quantities outside (inside) parenthesis, the values refer to 30 kpc (2$R_{\rm e,*}$) aperture. From left to right: subhalo identifier, stellar mass, total dark matter mass, stellar half-mass radius, maximum of the spherically-averaged rotation curve, mass-weighted age, mass-weighted stellar metallicity, [Mg/Fe] stellar relative abundance (solar units).}
\label{tab:properties}
\begin{tabular}{cccccccccc} \hline
Subhalo ID & log$(M_{\rm *}/\rm M_\odot)$ & log$(M_{\rm DM}/\rm M_\odot)$ & $R_{\rm e,*}$ & $V_{\rm max}$ & Age & log$(Z/Z_{\rm \odot})$ & [Mg/Fe] \\ 
  &  &  & [kpc] & [km s\textsuperscript{-1}] & [Gyr] &  &  & \\ \hline
6 & 10.98 (10.8) & 11.13 & 1.57 & 396.9 & 11.30 (11.39) & 0.33 (0.41) & 0.54 (0.54) \\
63875 & 10.66 (10.48) & 11.04 & 1.62 & 459.3 & 11.34 (11.71) & 0.26 (0.34) & 0.51 (0.51) \\
282780 & 10.57 (10.38) & 11.48 & 1.32 & 299.5 & 11.85 (12.22) & 0.27 (0.39) & 0.52 (0.51)\\
366407 & 11.21 (11.04) & 12.85 & 1.79 & 543.9 & 12.10 (12.26) & 0.34 (0.42) & 0.54 (0.54)\\
516760 & 10.81 (10.62) & 12.25 & 1.36 & 361.8 & 10.87 (11.15) & 0.29 (0.39) & 0.51 (0.51) \\ \hline
\end{tabular}
\end{table*}

\subsection{Degree of relicness}\label{sec:relicness}
Previous works have explained the properties of the relic candidates in terms of a degree of relicness, which is related to how far into the path, to become a typical nearby massive ETG, a compact relic has moved \citep{FerreMateu17, spiniello2021, Alamo-Martinez2021}. In the context of this work, it is a degree which is highest for relic candidates that had the most passive evolution in the last 9.5 Gyr, that is, have not significantly changed their stellar structure and populations since $z = 1.5$, either by star formation or accretion. Therefore, in order to quantify the relicness of relic analogue candidates in the TNG50, we define a score which captures essential features of relic galaxies. To achieve the highest score, a relic analogue candidate should simultaneously satisfy the following conditions:
\begin{enumerate}
    \item Do not display stellar populations younger than 3 Gyr (no recent star formation);
    \item Have $\geq$ 95 per cent of $M_*$ composed of stellar populations older than 9.5 Gyr (strong age uniformity);
    \item Have $R_{\rm e,*}$ below median if compared to galaxies of similar $M_*$ and have $\sigma_{\rm c} \geq$ 200~km~s\textsuperscript{-1} (structural similarity);
    \item Have super-solar metallicity and super-solar [Mg/Fe] elemental abundance (chemical similarity);
    \item Have $\gtrsim$ 75 per cent of $M_*$ composed of stellar populations older than 9.5 Gyr (weak age uniformity);
\end{enumerate}
Conditions (i) and (ii) capture features of an idealised relic galaxy, which is the absence of young stellar populations and strong uniformity in age. Conditions (iii) and (iv) capture important chemical and structural features compatible to what is expected from observations, with the sample presented in section \ref{sec:obs_sample} serving as an observational reference. And finally, condition (v) captures the weaker age uniformity in objects that would not necessarily be similar to real relic galaxies, but nonetheless, have great amounts of old stellar populations. The conditions listed above are hierarchical in the sense that (i) has the highest weight. Therefore, candidates which satisfy all conditions, receive score 5, candidates which satisfy all conditions except (i), receive score 4, those which satisfy only (iii), (iv) and (v) receive score 3, and so on, up to candidates which satisfy only condition (v), which receive score 1. Candidates which have score $\geq$ 4 are considered strong relic galaxy analogues throughout this work.

\subsection{Mock observations}
Since the current work only uses single band photometric data for the real galaxies, a fair juxtaposition would be to compare the light distribution of the relic analogue candidates with the real relic galaxies. In order to do that, we generate HST H-band mock observations using a Python pipeline which runs \textsc{SKIRT} \citep{SKIRT8,Camps2020SKIRTGrains}. As the candidates presented here have negligible SFR and low cold gas content, the steps necessary to generate the mocks are greatly simplified as there is no significant young stellar component or dusty interstellar medium.  Each stellar particle in the simulation is considered as a Single Stellar Population (SSP) with \textsc{GALEXEV} spectral energy distributions \citep{GALEXEV} based on its stellar mass, absolute metallicity and age. Since each stellar particle is representing an SSP, they are spatially modelled with a smoothing length of a truncated Gaussian emissivity profile equal to the distance to its 64th neighbour particle, instead of being considered as a point source \citep{Trayford2017OpticalSKIRT}. Then, we define a wavelength grid covering any spectral features we want to probe within the HST H-band filter response function and, for each wavelength bin, $10^7$ photon packets are launched assuming isotropic emission until they reach the virtual detector.
\par For each relic candidate, we produce observations in three different orientations aligned with the simulation box axis xy, xz and yz. {\sc SKIRT} outputs a data-cube with a SED for each pixel. HST H-band broadband images are then generated by convolving the data-cube with the filter response function of the HST H-band. This produces a noise-less image which is then matched to HST observational features by properly convolving it with the telescope point spread function and assigning correct noise levels.

\subsection{Photometric methods}\label{sec:morph_methods}
To build the brightness profiles from the H-band images, we measure intensities inside elliptical apertures and annuli. First we use \textsc{sextractor} \citep{Bertin96} to generate segmentation maps, from these maps we construct the masks simply by removing the galaxy segmentation in each image. We identify the centre of the galaxies as the brightest pixel inside their segmentation and use it as a first guess for the fitting of elliptical isophotes with \textsc{photutils} isophotal analysis routines \citep{bradley2020}. Once the models capture the overall brightness distribution in the data, we estimate the mean ellipticity ($\epsilon$) and position angle (PA) from the model elliptical isophotes, excluding the innermost and outermost ellipses. Then, we perform simple aperture photometry using elliptical apertures with fixed $\epsilon$ and PA, building brightness profiles and computing mean and total intensities for different semi-major axes. 

\subsection{Potential issues when comparing simulations and observations}
One has to be careful when comparing properties of simulated galaxies with those of real galaxies, either because the quantities in the simulation do not match  real observables or because scaling relations in the simulations are shifted with respect to the their observational counterparts \citep{Donnari2019,Popping2019}. In the case of this work, the lack of a large samples of observed and simulated relics does not allow us to perform a robust comparison between the two populations, and both relic samples could be affected by different selection biases. 
In the next section, we will frequently compare the properties of relic analogue candidates to other galaxies in the TNG50 simulation. We note that this is done loosely, comparing with the face values of real galaxies. In this way, we can investigate if the candidates deviate from the overall population of simulated quiescent galaxies in the same sense as the observed relic galaxies deviate, e.g. relic galaxies seem to be $\alpha$-enhanced systems, so simulated analogues should have $\alpha-$enhancement with respect to the median values of other simulated quiescent galaxies. In fact, we talk more in terms of what is \textit{expected} from the observational constraints, checking and discussing how much the analogue candidates satisfy such expectations. 
\par Although TNG50-1 run has a high resolution, having stellar softening length ($\varepsilon_{\rm *}$) of 0.288 kpc, we still have to be careful when analysing stellar structures inside 2.8$\varepsilon_{\rm *}$. Such caution is justified because this is the minimum separation between particles below which the gravitational forces are Newtonian \citep{Pillepich18b}. Another important aspect for the conclusions drawn in this work is the volume of the simulation. Since TNG50 simulates a box with only $\sim$ 50 cMpc in size, it has an intrinsic statistical limitation regarding the number of rare objects that we can find.

\section{Results}\label{sec:Results}
In this section, we discuss the properties and the cosmic history of the 5 relic analogue candidates, comparing them to other quiescent galaxies in TNG50 and what is expected from the observational constraints, regarding their metallicity, $\alpha$-enhancement, age, stellar structure and kinematics. 

\subsection{Global properties and profiles}\label{sec:global_properties}
Here, we describe the properties of the objects in our final sample of 5 relic analogue candidates, comparing them with other quiescent galaxies from TNG50 and the constraints imposed by observations. We briefly discuss many global and spatially-resolved properties of the subhalos, in order to have a comprehensive characterisation for each candidate. The main goal of this section is to check which subhalos have characteristics in accordance to relics. Relevant properties of the 5 relic analogue candidates are presented in Table \ref{tab:properties}. Global properties - such as $M_{\rm *}$, metallicity and [Mg/Fe] - presented as scalar values in Fig. \ref{fig:properties}, are calculated within a spherical aperture of 30 kpc of radius\footnote{We also use a spherical aperture with radius equal to 2$R_{\rm e,*}$, but only for test purposes, since the results and conclusions presented throughout this work remain unchanged independently of which of the the two apertures is used.}.

\subsubsection{Chemical properties and age}
\begin{figure*}
    \centering
    \includegraphics[width=0.97\textwidth]{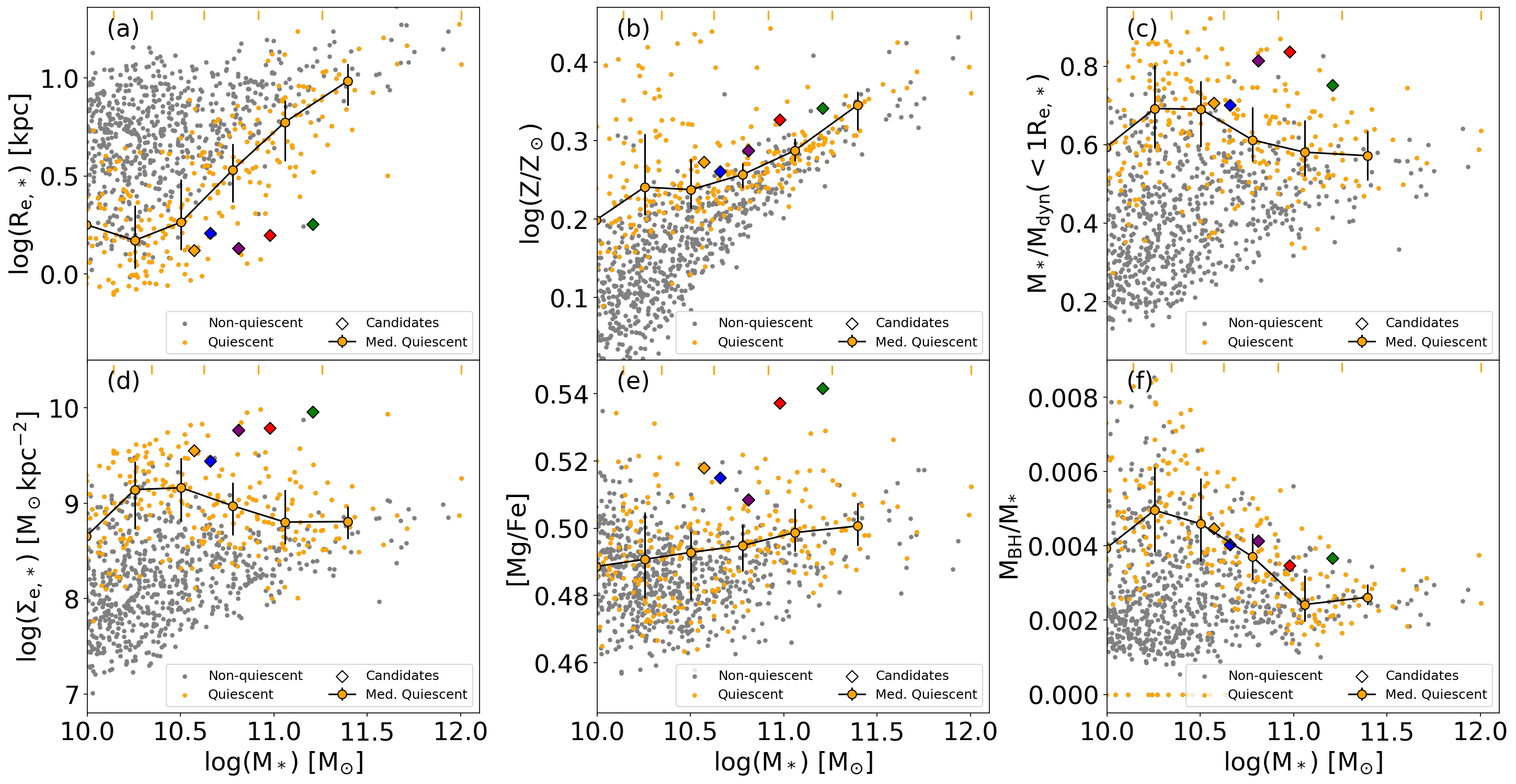}
    \caption{Global properties of subhalos in TNG50 versus their stellar mass (in 30 pkpc aperture). \textbf{Panel (a):} half-mass radius ($R_{\rm e,*}$). \textbf{Panel (b):} stellar metallicity ($Z$). \textbf{Panel (c):} stellar-to-dynamical mass ratio ($M_{\rm *}/M_{\rm dyn}$) inside 1 $R_{\rm e,*}$. \textbf{Panel (d):} stellar surface density inside 1 $R_{\rm e,*}$ ($\Sigma_{\rm e,*}$). \textbf{Panel (e):} [Mg/Fe] elemental abundance. \textbf{Panel (f):} black hole-to-stellar mass ratio. Quiescent subhalos are shown as orange circles, non-quiescent subhalos are shown as gray circles, relic analogue candidates are shown as coloured diamonds. Median values for the global properties of the quiescent subhalos in each stellar mass bin are shown as orange circles connected by black solid lines, 25th and 75th percentiles are indicated by the bars. The stellar mass bins edges are represented by the orange ticks at the top of the panels, in log$(M_{\rm *})$: 10.14, 10.35, 10.63, 10.92, 11.26, 12. The edges are chosen so that there are at least 50 data points in each bin, except for the higher mass bin, which has a lower number of points.}
    \label{fig:properties}
\end{figure*}

\begin{figure}
    \centering
    \includegraphics[width=0.47\textwidth]{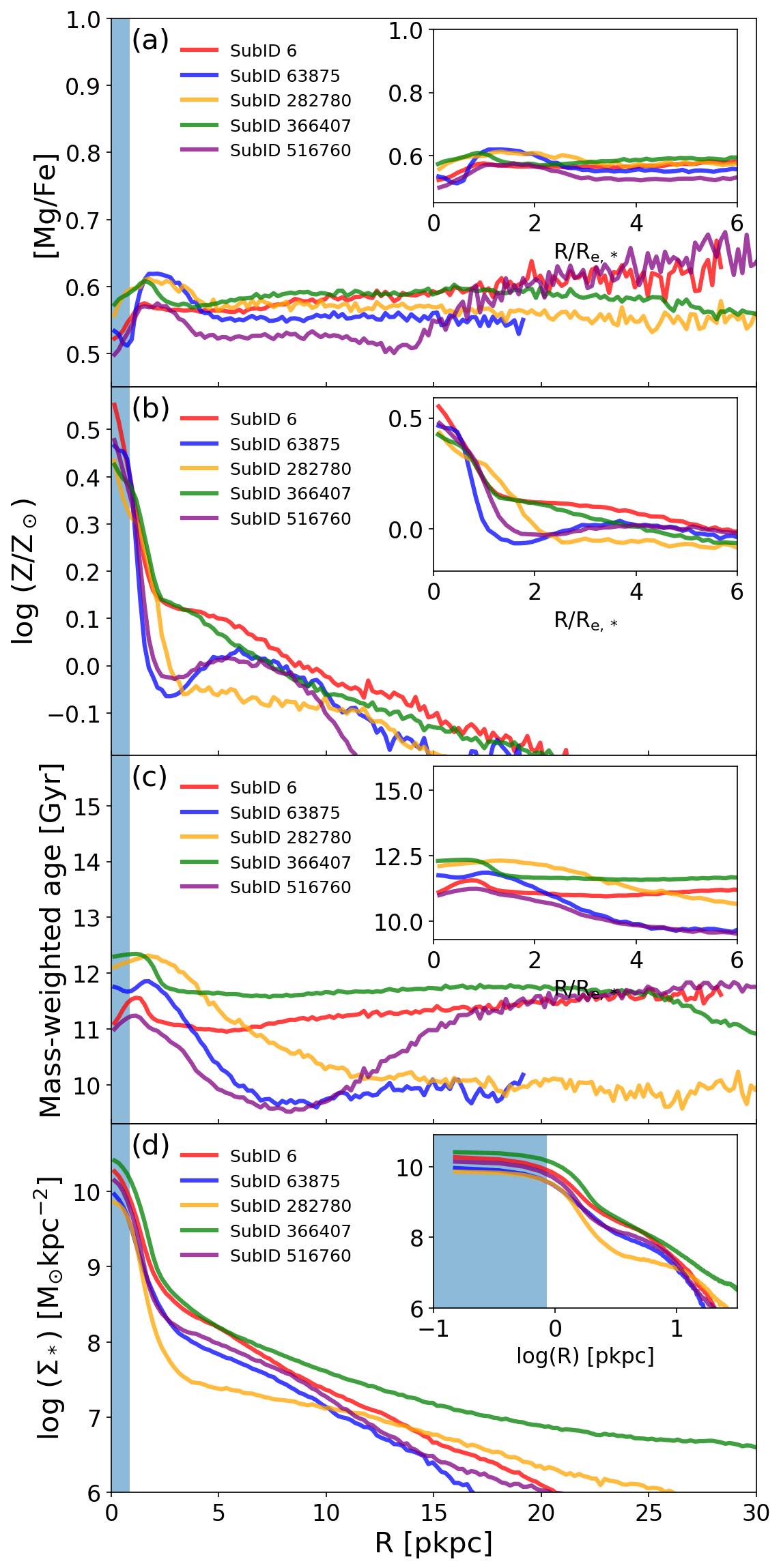}
    \caption{Radial distribution of different properties of the relic analogue candidates. \textbf{Panel (a)}: Stellar [Mg/Fe] elemental abundance, in solar units. \textbf{Panel (b)}: stellar metallicity, in solar units. \textbf{Panel (c)}: Mass-weighted stellar age. \textbf{Panel (d)}: Stellar surface density. For details regarding the estimate of each profile, see section \ref{sec:Methods}.}
    \label{fig:profiles}
\end{figure}

In panel (b) of Fig. \ref{fig:properties}, we plot mass-weighted stellar metallicity vs stellar mass of relic candidates, also comparing with other quiescent galaxies in the simulation. As we can see in this panel, the candidates are metal-rich objects which lie above the median values for quiescent galaxies of similar $M_{\rm *}$, with subhalos 6 and 366407 being the most metal-rich systems among the candidates. As shown in Table \ref{tab:obs_properties}, relic galaxies from Y17 sample have super-solar metallicities, similar to what is observed in the central parts of the most massive and oldest nearby ETGs \citep{Gallazzi2005,Panter2008,GonzalezDelgado2014}, thus, the super-solar metallicities in the relic analogue candidates agrees with what is expected from the observational constraints. 
In panel (b) of Fig. \ref{fig:profiles}, we present metallicity profiles for all the candidates. We can see that the metallicity of these subhalos decreases significantly from the centre to regions at $R>1R_{\rm e,*}$, nonetheless, most of them exhibit super-solar mass-weighted metallicities inside 2$R_{\rm e,*}$, as is expected from the properties of observed relics, which have metal-rich populations in their central parts \citep{FerreMateu17,yildirim17}. We note that Subhalo 6 and Subhalo 366407 remain super-solar out to 4$R_{\rm e,*}$ ($\sim$~10~kpc), in better agreement to what is expected from metallicity profiles of the observational sample considered here (check Figure 10 in Y17). 

\par A further interesting analysis into the chemical properties of the candidates, is the stellar [Mg/Fe] elemental abundance, which can be used as a proxy for $\alpha$-enhancement. We can see in panel (e) of Fig. \ref{fig:properties} that the relic analogue candidates are among the most $\alpha$-enhanced subhalos for their respective stellar masses, with subhalos 6 and 366407 being the most extreme cases among the candidates. The spatial distribution of [Mg/Fe] throughout the subhalos - shown in panel (a) of Fig. \ref{fig:profiles} - seems to be nearly uniform, presenting typical values around 0.55 dex, with the most extreme radial gradients located in the region enclosed by 1$R_{\rm e,*}$. The relatively high values of [Mg/Fe] of the candidates are in accordance with these objects forming the majority of their stellar content in a shorter time-scale than other quiescent subhalos. In general, the global metallicity and $\alpha$-enhancement for most of the relic analogue candidates are in agreement with what is expected from the observations \citep{FerreMateu17,yildirim17}. However, the spatial distribution of their chemical properties do not agree so well with observations. The simulated galaxies have  more abrupt changes in their metallicity profiles inside 2 $R_{\rm e,*}$, in some cases reaching $\Delta \log(Z) \sim 0.4$ dex (see Fig. \ref{fig:profiles}). Nevertheless, we note that subhalos 6 and 366407 present both roughly uniform [Mg/Fe] profiles and monotonic decreasing super-solar metallicity out to 10 kpc, qualitatively similar to the real relic galaxies considered as reference in this work. 

Another important feature in observed relic galaxies is their age profiles, which have been used during the sample selection  of this work (see section \ref{sec:sample_selection}). The only constraint during the selection was that all the selected subhalos should have age profiles with values above 9.5 Gyr. Here we analyse the shape of such profiles. In panel (c) of Fig. \ref{fig:profiles}, we present mass-weighted age profiles of the candidates. Despite the fact that these simulated galaxies are quite old up to 30 kpc, some of them have age differences within their profiles that can reach 2 Gyr. Subhalo 6 and Subhalo 366407 are the most uniformly old objects, showing maximum age differences smaller than 500 Myr throughout their profiles. This near uniformity in age is another evidence in favour of these 2 objects being strong relic analogues. We have also computed the age profiles using $r$-band luminosity weights, but there was no significant difference in the results.  

\subsubsection{Stellar component structure}\label{results:stellar_struct}
It is expected that any relic analogue in the simulation will be compact when compared with galaxies of similar stellar mass. In panels (a) and (d) of Fig. \ref{fig:properties}, we compare the size and stellar surface density of the candidates with other galaxies in the simulation. As expected from the selection criteria employed in section \ref{sec:sample_selection}, relic analogue candidates are all below the median size of quiescent galaxies in the $M_{\rm *}-R_{\rm e,*}$ diagram. In particular, the most massive candidates (subhalos 6 and 366407) are almost 1 dex away from the median values of their bins, being very compact for their respective stellar masses. The compactness of these two objects is also evident by their effective stellar surface densities ($\Sigma_{\rm e,*}$), shown in panel (d) of Fig. \ref{fig:properties}. 

In order to analyse the internal structure of the candidates, we also construct stellar surface density profiles ($\Sigma_{\rm *}(R)$) of the subhalos, which are shown in panel (d) of Fig. \ref{fig:profiles}. One noticeable feature present in all candidates is a core region inside $R \leq$ 0.8 kpc with nearly constant density. This is in contrast with the structure of observed relic galaxies at $z \sim 0$ and quiescent galaxies at $z > 0.5$, which have a steeper surface density profile in their innermost parts, as can be seen in \cite{yildirim17} (their Figure 6). Such regions of nearly constant density are within the 3$\varepsilon_{\rm *}$ limit, where numerical effects may be important. There is a very high number of stellar particles inside this region and at the same time, gravitational forces become Newtonian only at a separation of 2.8$\varepsilon_{\rm *}$ \citep{Pillepich2018a}. Therefore, we speculate that the high number of particles inside a potential which is not Newtonian could generate artificial structures that are not observed in real compact galaxies. This core region is also present in their brightness profiles (section \ref{results:morph}).
Although the density profiles of the relic analogue candidates seem to be discrepant from observations in their central regions, the outer parts of their stellar surface density profiles are comparable with the values observed in the Y17 galaxies (see their Figure 6), going from $\Sigma_*(1 $kpc$) \sim 10^{9.6}$ $\rm M_\odot kpc^{-2}$ to $\Sigma_*(\sim 20 $kpc$) \sim 10^{6}$ $\rm M_\odot kpc^{-2}$. The density profiles of the relic analogue candidates also present visible changes in their slopes, suggesting that these subhalos have different structural components, which is reinforced by the light distributions and kinematic maps presented in the next subsections. 

\subsubsection{Mass fractions}
The dynamical analysis done by Y17 showed that the compact elliptical galaxies used in their work have high stellar-to-dynamical mass ratios $M_{\rm *}/M_{\rm dyn}$ in the inner regions, with a mean value of $M_{\rm *}/M_{\rm dyn} (R<R_{\rm e,ell})$ = 0.89 for their whole sample. In the simulation, we measured $M_{\rm *}/M_{\rm dyn}$ inside 1$R_{\rm e,*}$ considering $M_{\rm dyn}$ as the total mass of all types of particles, and we show the results in panel (c) of Fig. \ref{fig:properties}. We can see that the relic analogue candidates have higher $ M_{\rm *}/M_{\rm dyn}$ than other quiescent subhalos of similar $M_{\rm *}$, with subhalos 6, 366407 and 516760 having the highest $M_{\rm *}/M_{\rm dyn}$ values among the candidates, in agreement to what is expected from the observational sample. The dark matter mass fractions inside 1$R_{\rm e,*}$ ($f_{\rm DM}$) of these 3 candidates are 15.7\%, 24.4\% and 17.9\% respectively, while the median $f_{\rm DM}$ of Y17 sample is only $\sim 11$\% inside 1 half-light radius. In order to draw more robust conclusions about these mass fractions, a more complete and detailed dynamical analysis is needed, allowing to compare more directly $M_{\rm *}/M_{\rm dyn}$ and $f_{\rm DM}$ between observations and simulations. However, given the complexity of this dynamical modelling, this analysis is left to a follow-up study. 
Additionally, Subhalo 6 is special among the relic analogue candidates because of its very low total dark matter-to-stellar mass fraction, which reaches log$(M_{\rm DM}/M_{\rm *})$ = 0.15. As discussed further in section \ref{sec:cosmic_evo}, this low dark matter fraction is a result of the interaction of this object with its current galaxy cluster.
\par Another interesting property to check in the  relic analogue candidates is their black hole (BH) masses. Panel (f) of Fig. \ref{fig:properties} shows that central BH masses of the TNG analogues are not much distant from the median value for quiescent galaxies of similar mass in the simulation, in contrast with the observational sample of Y17. However, it is important to note that the Y17 sample is biased towards galaxies with "over-massive" black holes, and maybe this can contribute to the discrepancy. Additionally, given the immense dynamical scale involved in BH growth and feedback, this kind of modelling still remains a challenge for cosmological simulations \citep{Vogelsberger2020,Hopkins2021}. Thus, it can potentially contribute to discrepancies between the properties of simulated and real relic galaxies, e.g. the stellar density in their central regions. 

\subsection{Light distribution}\label{results:morph}
\par Besides the distribution of mass traced by the stellar particles in the simulation, it is also relevant to address the structure of the relic analogue candidates using their projected light distributions. It is important to check whether their overall morphologies are compatible to those of observed relics. Here we are taking a step further in the comparison of the simulation with observations, creating mock observations of the candidates in a specific photometric filter (HST-F160W) and analysing if their light distributions are qualitatively similar to the ones of real galaxies from Y17 - also observed with the filter F160W (H-band) of HST. 
\par
\begin{figure*}
    \centering
    \includegraphics[width=0.98\textwidth]{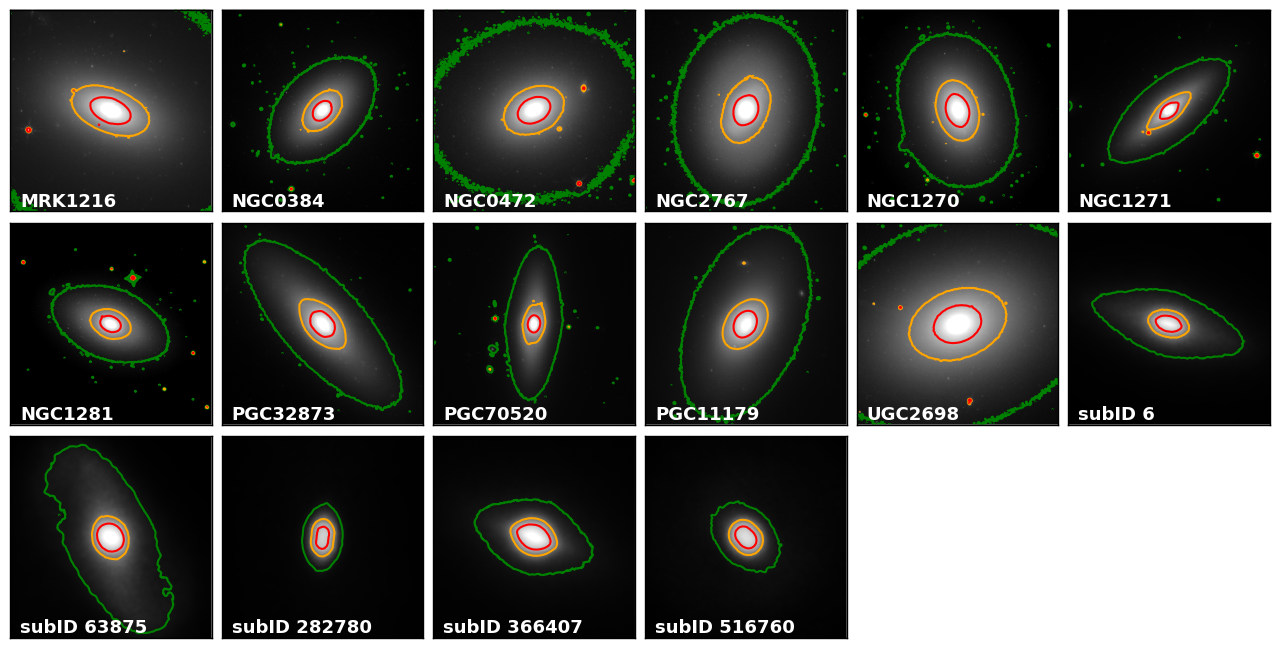}
    \caption{H-band images of both observed and simulated galaxies analysed in this work. Simulated galaxies are identified by their subhalo ID's. All images are normalised by the intensity at the half-light radius of each galaxy ($I_{\rm e}$). They have the same scale (asinh) and approximately the same FOV (20 kpc $\times$ 20 kpc). Each isophote represents fractional values of $I_{\rm e}$, from the outermost to the innermost isophotes, the fractions are: 0.1 (green), 1.0 (orange), 3.0 (red). For more details, see section \ref{sec:obs_sample}.}
    \label{fig:H-band}
\end{figure*}

\begin{figure}
    \centering
    \includegraphics[width=0.47\textwidth]{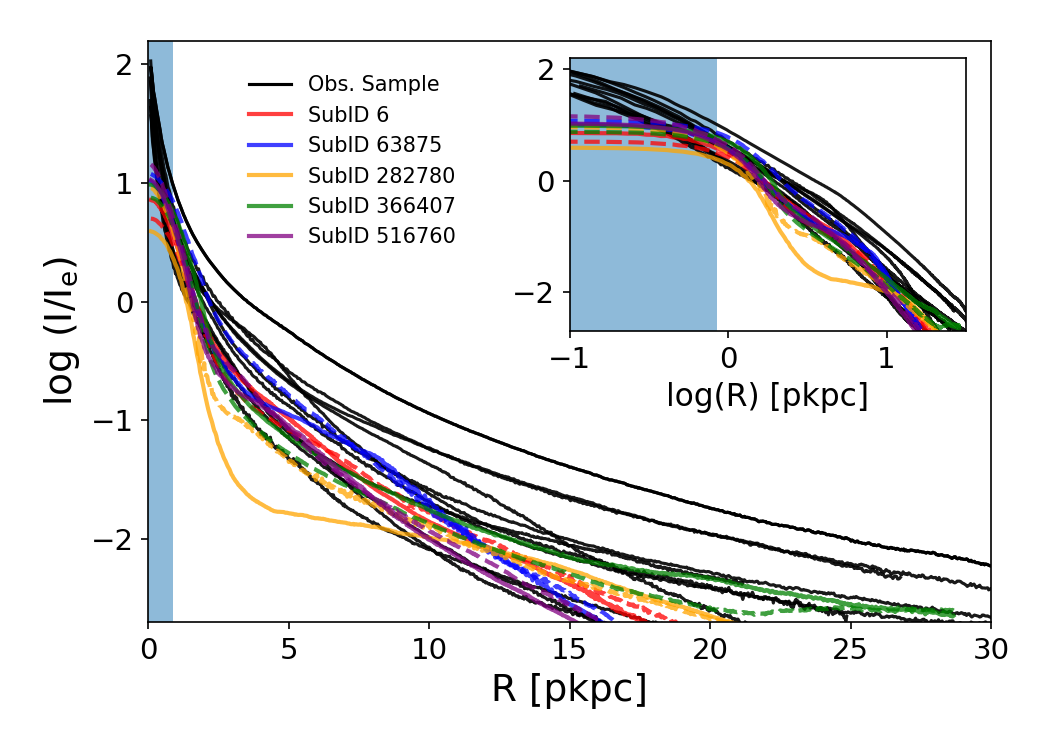}
    \caption{Brightness profiles for the relic analogue candidates in TNG50 and galaxies from the observational sample. The intensities are normalised by the intensity at the effective radius. The blue shaded stripe indicates the region inside 3 softening lengths. Dashed and solid lines indicate 2 different random orientations for the same galaxy. The inset is the same plot with radial axis in log scale, to highlight the profiles in the inner regions of the subhalos.}
    \label{fig:profile_light}
\end{figure}

 In Fig. \ref{fig:H-band}, we show the H-band images for the real and simulated galaxies used in this work. From the images we can qualitatively analyse the shape of the isophotes\footnote{The value for each isophote is actually the brightness ratio $I/I_{\rm e}$, where $I_{\rm e}$ is the intensity at the half-light radius.} and notice two important features. First, the simulated galaxies appear to have shallower brightness profiles in their cores. As we can see in Fig. \ref{fig:H-band}, the inner isophotes of the simulated galaxies enclose a smaller area than in the real ones, meaning that in the centre, they reach relatively smaller fractional values of their $I_{\rm e}$. Second, both observed and simulated samples have galaxies with disky isophotes, having morphologies that resemble the S0 visual classification in some cases (e.g. NGC 1271, PGC 32873 PGC 70520, Subhalo 6 and Subhalo 366407). The first feature arises from the underlying discrepancy between the inner stellar surface densities of real relic galaxies and the simulated galaxies, as discussed in Subsection \ref{results:stellar_struct}.
The results presented in Fig. \ref{fig:profile_light} show that the brightness profile shape of the simulated galaxies is different from the shape of real relics at $R < 1$~kpc. Similarly to the surface density profile, there is a region inside $R < 0.8$~kpc which has nearly constant brightness levels.

Regarding more detailed descriptions of the morphology, we can see in Fig. \ref{fig:H-band} that both real and simulated galaxies do not have perfectly elliptical isophotes with constant $\epsilon$ and PA. In fact, some of the real galaxies used as observational constraints in this work seem to have more complex structures. For example, galaxies NGC 1270, NGC 0374, NGC 0472 and NGC 2767 were photometrically decomposed by \cite{Lucatelli2019} based on the curvature of their light profiles, showing evidence for substructures - such as disks - in their inner regions. The simpler analysis done here already shows the two most important aspects regarding the light distribution: overall morphological resemblance between some subhalos in the simulation and some of the real relics, and discrepant light profile slopes in the inner regions ($R < 1$~kpc) of all simulated galaxies.

\subsection{Stellar kinematics}\label{results:kinematics}
\begin{figure*}
    \centering
    \includegraphics[width=0.96\textwidth]{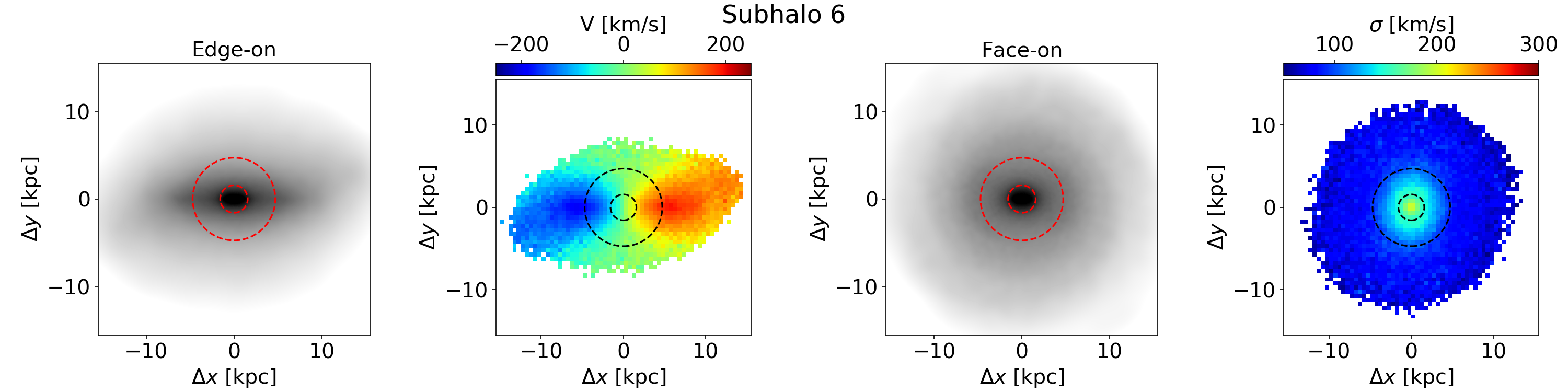}
    \includegraphics[width=0.96\textwidth]{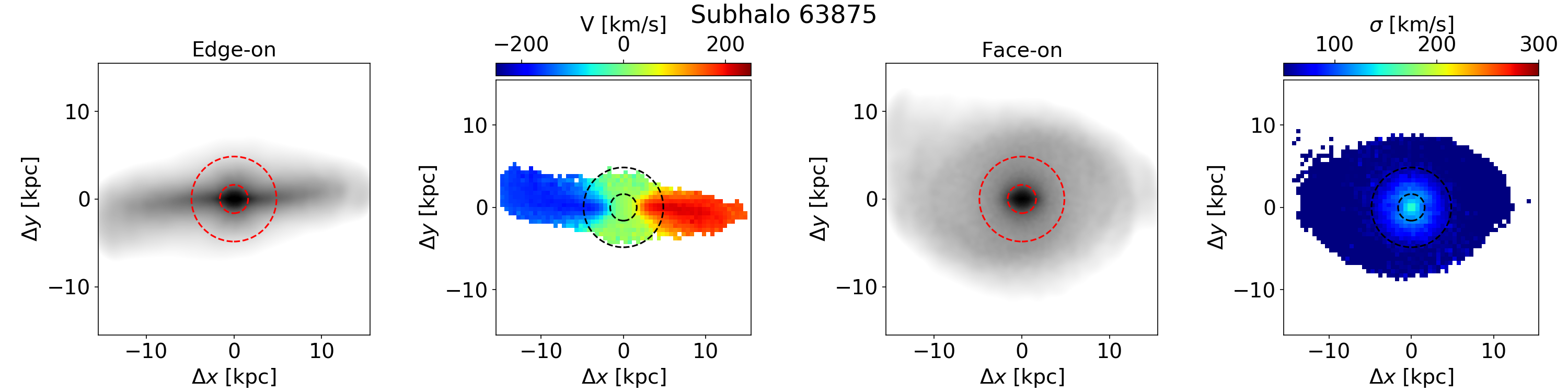}
    \includegraphics[width=0.96\textwidth]{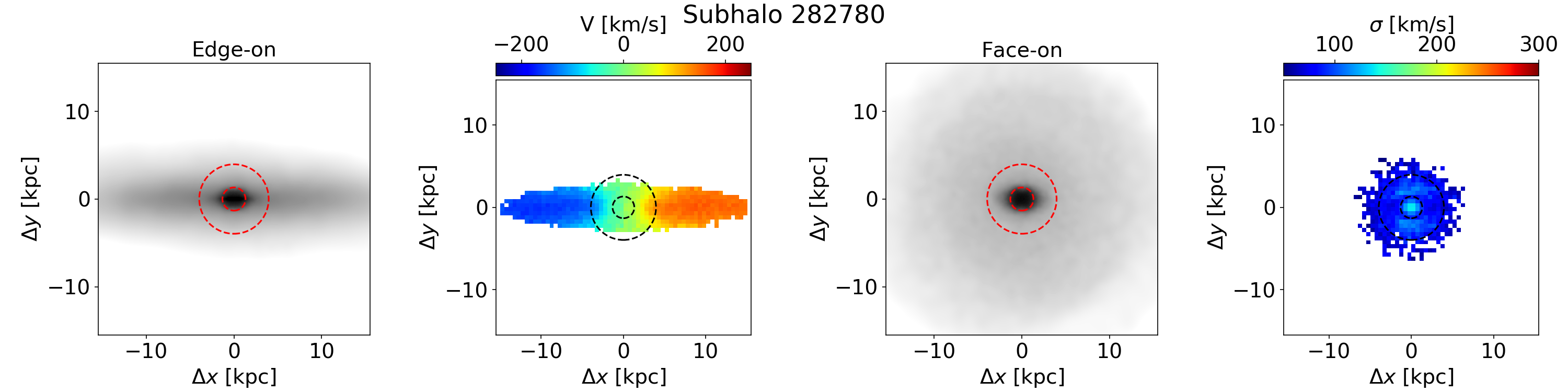}
    \includegraphics[width=0.96\textwidth]{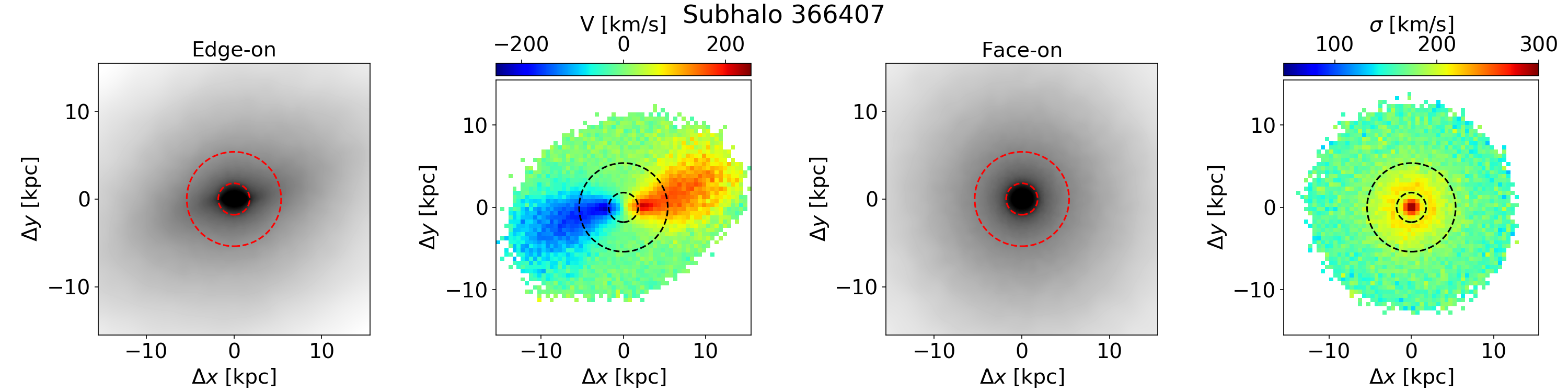}
    \includegraphics[width=0.96\textwidth]{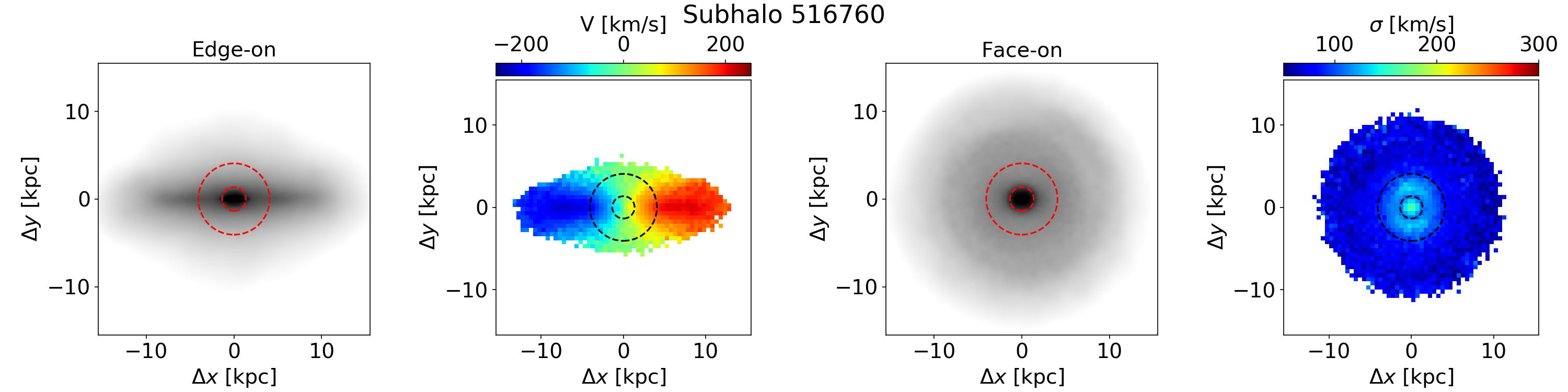}
    \caption{Mass-weighted kinematic maps constructed from stellar particles using square pixels 0.5 kpc wide. Velocity maps are constructed from the edge-on orientation, while velocity dispersion maps are constructed from the face-on orientation (for random orientations check the Supplementary Material). Dashed circles indicate 1 and 3 $R_{\rm e,*}$. To the left of each map, we show the respective stellar surface density for the same orientation.}
    \label{fig:kinematics}
\end{figure*}

Another approach to study the structure of galaxies is to analyse their kinematic maps. In this subsection, we present some kinematic maps that can give us a glimpse on the orbital structures of the relic analogue candidates of TNG50. In the velocity maps of Fig. \ref{fig:kinematics}, all the candidates present evidence of a rotation disk, which we could already infer from their apparent S0-like morphologies (see Fig. \ref{fig:H-band} in section \ref{results:morph}). They also present a region with relatively high velocity dispersion within 1 $R_{\rm e,*}$ and, in general, the central velocity dispersion ($\sigma_{\rm c}$) observed in the maps can be greater if we use higher spatial resolutions or random line-of-sights (see maps in Supplementary Material). Therefore, the maximum $\sigma$ observed in the privileged face-on projections of Fig. \ref{fig:kinematics} are considered here as minimum values for $\sigma_{\rm c}$ of these subhalos. Additionally, if the gravitational softening artificially decreases the central density of subhalos, it may also reduce the velocity of stellar particles in the centre, possibly underestimating $\sigma_c$ of the subhalos in comparison to real galaxies.

The higher values of $\sigma$ are observed in the centre of Subhalo 6 and Subhalo 366407, reaching $\sim 200$~km~s\textsuperscript{-1} and $\sim 300$~km~s\textsuperscript{-1} respectively. Although Subhalo 6 reaches this high $\sigma_{\rm c}$, it is still a relatively low velocity dispersion if directly compared with $\sigma_{\rm c}$ face values of observed relic galaxies (see Table \ref{tab:obs_properties}). Since the kinematic maps presented in this work are not the derived from mock integral field spectroscopy observations, comparison between absolute values of velocities must be taken carefully. Considering this limitation and analysing the available kinematic maps in Fig. \ref{fig:kinematics}, we conclude that subhalos 6 and 366407 are the only candidates which have $\sigma_{\rm c}$ minimally comparable to the observations, at least by order of magnitude. Nevertheless, it is also important to note that the Y17 sample is biased towards objects with high central velocity dispersions (see Y17 Introduction). 
\par Regarding the rotational support, all galaxies in our observational sample are classified as fast rotators by Y17 (following the classification scheme of \cite{Emsellem2011}), although UGC 2698 is only at the edge of being classified as fast rotator (see Y17 Discussion). Further kinematic analysis would be required in order to confirm whether the relic analogue candidates presented here could be classified as fast rotators according to some commonly used criteria in observational studies. Nevertheless, we can expect that the simulated galaxies in TNG50, even the fast rotators, will have lower rotation-to-dispersion ($V/\sigma$) values if compared to observations. This expectation is supported by a recent work of \cite{Pulsoni2020}, which showed that fast rotators ETGs in TNG50 and TNG100 have shallower $V/\sigma$ profiles than real ETGs, with many of the real ETGs that they compared, being from ATLAS\textsuperscript{3D} \citep{Capellari2011}. \cite{Pulsoni2020} show that very few fast rotators in the simulation reach $V/\sigma \geq 1$ within 1 $R_{\rm e,*}$ and suggest that the angular momentum distribution in the simulations could be different from the real galaxies.
\par We do not attempt any detailed photometric or kinematic structure decomposition in this work, but the kinematic results shown here suggest that the relic analogue candidates in our final sample have multiple components, in accordance with previous results already presented in \ref{results:stellar_struct} and \ref{results:morph}. Our relic analogue candidates have dense cores with radius of $\sim$ 1 kpc, which are dominated by dispersion, and more extended ($R > 3$~kpc) well-defined rotation disks. Such structures are still present even if we change the resolution of the maps\footnote{We encourage the reader to check kinematic maps with higher resolution in the Supplementary Material available online.}. The clear rotational patterns and high central velocity dispersions present in Subhalos 6 and 366407 are in qualitative agreement with the kinematic maps of the real relics (see Appendix B of Y17), thus, supporting the hypothesis that these two objects represent strong relic analogues.

\subsection{Cosmic history}\label{sec:cosmic_evo}
So far we have investigated several properties of the relic analogue candidates at $z=0$ and compared them with the expected properties of relic galaxies. Now we analyse in detail the cosmic history of each relic candidate in the simulation. In this section we aim to characterise the stellar assembly of the candidates, analysing in detail their size evolution, stellar \textit{in-situ} mass fractions and star formation history, often comparing their stellar assembly to what is expected for idealised relic galaxies (early and rapid stellar build-up followed by null or negligible growth). Also and equally important, we aim to study the evolution of gas and dark matter of the candidates, analysing how these components are affected by internal or external mechanisms. We suggest the reader to check the videos we produced using snapshot data, available online as Supplementary Material. Except when it is explicitly mentioned, stellar mass ($M_{\rm *}$) corresponds to a 30 kpc aperture.

\subsubsection{Subhalo 6}\label{sec:history_sub6}
\begin{figure*}
    \centering
    \includegraphics[width=0.9\textwidth]{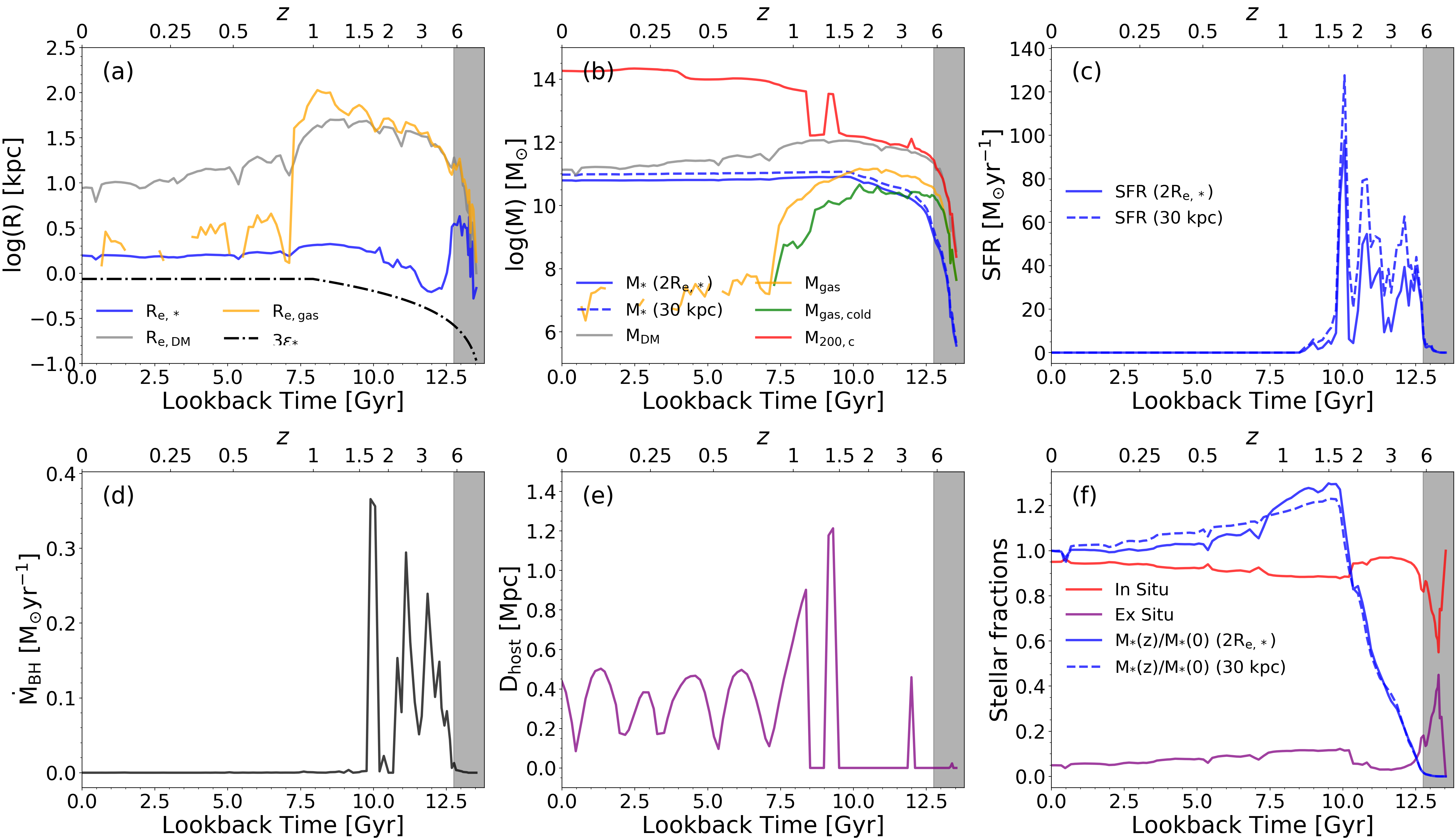}
    \caption{Cosmic history of Subhalo 6 represented by the evolution of many of its properties. \textbf{Panel (a):} stellar (blue), gas (orange) and dark matter (grey) half-mass radii, dot-dashed line represents 3 softening lengths. \textbf{Panel (b):} stellar (blue), gas (orange), cold gas (green), dark matter (grey) and host halo (red) masses. \textbf{Panel (c):} instantaneous SFR in different apertures. \textbf{Panel (d):} instantaneous BH accretion. \textbf{Panel (e):} Radial distance to centre of host halo. \textbf{Panel (f):} Stellar mass fractions with respect to $z=0$ in different apertures (blue), relative \textit{in-situ} (red) and \textit{ex-situ} (purple) fractions in each snapshot. In panel (b) the gas represented in orange is all gravitationally bound gas, while the cold gas represented in green is only the gas inside 30 pkpc which has temperature $<10^5$~K. Black shaded regions indicate the epochs in which the subhalos are not considered to be resolved. Blank intervals in a curve usually indicate that the quantity is not computed in a particular snapshot, or in case of logarithmic axes, the quantity had a zero value and could not be represent in the plot.}
    \label{fig:history_6}
\end{figure*}

In Fig. \ref{fig:history_6}, we present the cosmic evolution for many properties of Subhalo 6. This subhalo already formed 80 per cent of its stellar content at $z \sim 1.9$, mostly through in-situ star formation, as can be seen in panels (c) and (f) of Fig. \ref{fig:history_6}. At $z \sim 1.7$ the subhalo already assembled 100 per cent of its current stellar mass, after a gas rich minor merger at $z \sim 1.75$ triggered the last burst of star formation (panel c), which was accompanied by activity in the central black hole (panel d). This merger slightly increased the ex-situ fraction (panel f) by up to $\sim 10$ per cent, as expected. Soon after, at $z \sim 1.2$ (8.7 Gyr ago) this relic analogue candidate experiences a change in environment, entering in the progenitor halo of the most massive cluster ($M_{\rm 200,c} > 10^{14} \rm M_\odot$) of the simulation at $z=0$. Subhalo 6 remains a satellite of this cluster since then and during the infall, it gradually looses dark matter, having  log$(M_{\rm DM}/M_{\rm *}) = 0.15$ at $z=0$. Recent work by \cite{Engler2020} showed that satellite galaxies in TNG simulations suffer tidal stripping of their dark matter structure as they enter in their host halo, regardless of their stellar mass (check their section 4). This seems to be the case for Subhalo 6 (see panel e in Fig. \ref{fig:history_6}), which already performed 5 pericentric passages since $z = 1.2$ and lost $\sim 90$ per cent of its dark matter content (see panel b in Fig. \ref{fig:history_6}). Moreover, almost all remaining gas at $z \sim$ 1.2 is removed from this subhalo during the first approach to the host halo centre (Fig. \ref{fig:history_6}, panels a, b and e). Before entering the cluster, Subhalo 6 has a well-defined gas disk, which is not present anymore at $z \sim 1.2$. By visually analysing the complex gas dynamics around this galaxy we notice that the gas disk is apparently destroyed between $z=1.3$ and $z=1.2$. We see that it does not have any significant BH accretion around the time the disk is destroyed ($\sim$ 8.8 Gyr ago) nor is close to the centre of its current host halo, as shown in panel (e). We suggest that the destruction was caused mainly due to instabilities in the gas introduced by the close passage of neighbouring galaxies. However, further analysis is needed in order to tell exactly what destroyed the gas disk. Nonetheless, the very early intense consumption and subsequent loss of gas prevented any recent star formation in Subhalo 6, contributing for is uniformly old age at $z = 0$.
\par During the infall into the cluster, the stellar size was moderately affected, the galaxy shrunk $\sim 0.4$ kpc and lost $\sim 19$ per cent of its stellar content (see panels (a) and (f) in Fig. \ref{fig:history_6}). Given that the dark matter content of this subhalo may have been considerably stripped, we suggest that the stellar component has diminished also due to tidal stripping. Additionally, the \textit{in-situ} stellar component of this subhalo grew by $\sim 5$ per cent during the loss of its $M_{\rm *}$, implying that more of the original \textit{in-situ} component survived, relatively to the \textit{ex-situ} one. Both components were assembled prior to $z \sim 1.7$, so the ex-situ component does not represent a population of younger stars, just something that came from an early merger at $z = 1.75$.
Although the stellar mass assembly of Subhalo 6 occurred very early ($z>1.5$) and that it is still a very compact and old object, we cannot neglect the interaction of this simulated galaxy with its parent cluster. In other words, we cannot claim this is a \textit{completely untouched} relic analogue, as it would be the case for an idealised relic. Nonetheless, the overall star formation history of this object is evidence in support of it representing a strong relic analogue.
\par It is important to note, that the decrease in mass of the subhalo components, specially in the outskirts ($R > 30$~kpc), can possibly happen artificially due to the method employed to identify substructures in the simulation (see section \ref{sec:Simulation} for a very brief description). As a satellite subhalo delves deeper into its parent halo, the dark matter or gas on its outskirts could be assigned to the parent halo, decreasing the size and mass of the subhalo as it moves inward, regardless of whether or not the outer parts have been \textit{physically} stripped away. This is an important caveat that may introduce an ambiguity for gas and dark matter stripping in Subhalo 6 and other satellite subhalos analysed in this section.

\subsubsection{Subhalo 63875}
\begin{figure*}
    \centering
    \includegraphics[width=0.9\textwidth]{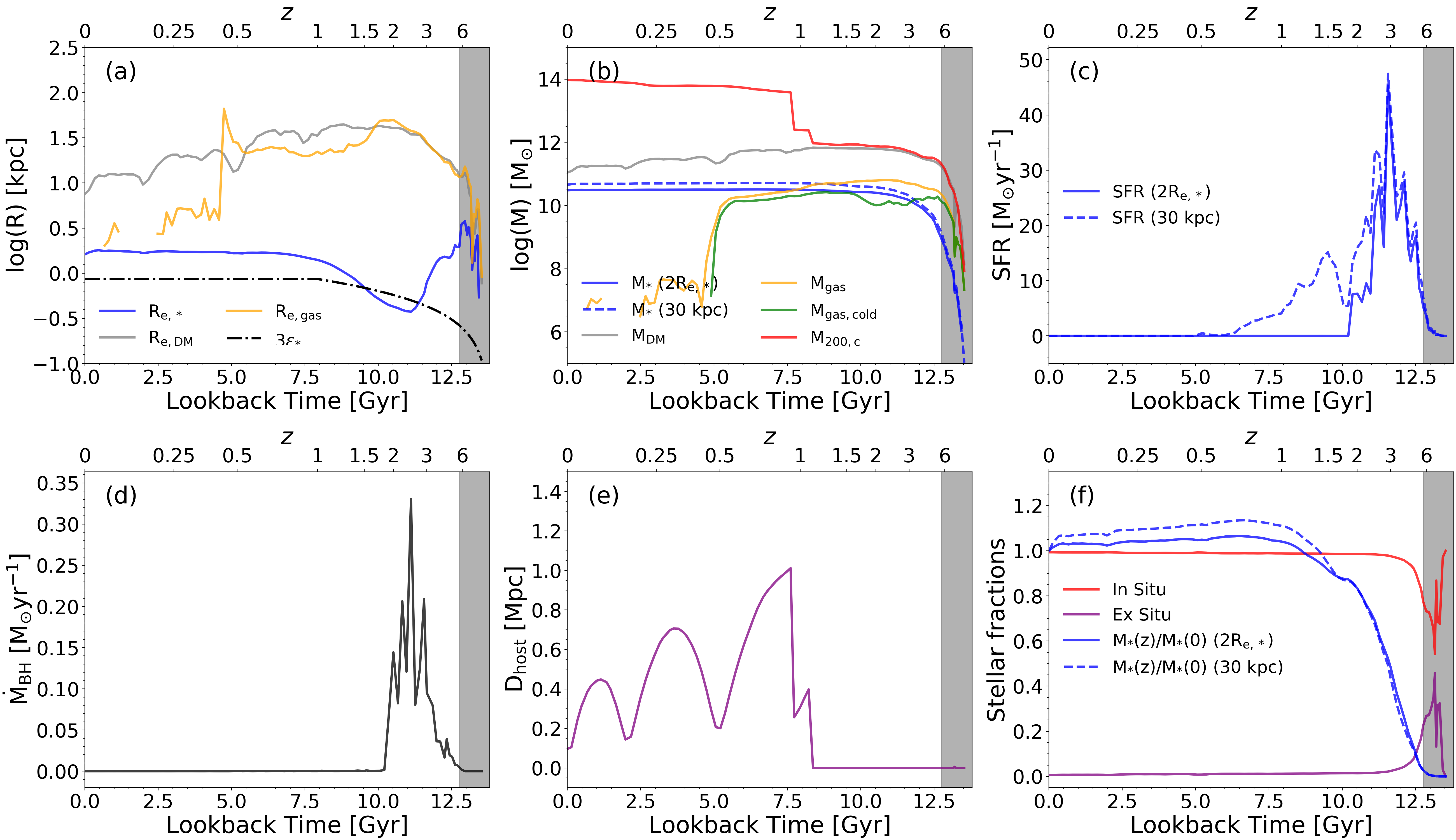}
    \caption{Cosmic history of Subhalo 63875 represented by the evolution of many of its properties. Same as in Fig. \ref{fig:history_6}.}
    \label{fig:history_63875}
\end{figure*}

In Fig. \ref{fig:history_63875}, we present the cosmic evolution for many properties of Subhalo 63875, as in Fig. \ref{fig:history_6}. The star formation of this subhalo can be divided into two epochs (panel c). In the first epoch, which lasted until $z \sim 1.8$, $\sim 87$ per cent of the current stellar mass of the subhalo was formed, almost $100$ per cent \textit{in-situ}, as shown in panel (f). The second epoch consisted of star formation activity in the outer parts ($R > 3$~kpc) of the subhalo, a continuously declining star formation episode which lasted until $z \sim 0.6$. This second epoch helps to drive a decrease of the stellar age of this candidate, with populations younger than 9.5 Gyr contributing almost $19$ per cent to its $M_{\rm *}$ at $z = 0$ (see Appendix \ref{app:stell_pop}). Consequently, this second star formation epoch left an imprint in the age profile of Subhalo 63875 (see panel c in Fig. \ref{fig:profiles}), with stars at $R = 5$~kpc being almost 2 Gyr younger than in the centre. 
This subhalo is also a satellite in a massive cluster ($M_{\rm 200,c} \sim 10^{14}$ $\rm M_\odot$) and similarly to Subhalo 6, had $\sim 80$ per cent of its dark matter removed, having log$(M_{\rm DM}/M_{\rm *})$ = 0.384 at $z \sim 0$. This object entered its host halo at $z \sim 0.8$ and had a very well-defined gas disk at $z \sim 0.6$, which was completely destroyed after the subhalo made its first pericentric passage at $z \sim 0.45$ (see panel (e) of Fig. \ref{fig:history_63875} and videos in Supplementary Material). Thus, the stripping interaction with the cluster seems to be the main driver behind the gas supply removal in this simulated galaxy, even if we consider only the cold gas supply. Regarding the stellar size, the $R_{\rm e,*}$ of this subhalo remains nearly constant - $\sim$1.7 kpc since $z \sim 0.8$ - even after it lost $\sim 12$ per cent of its stellar mass during the infall. 
\par Although Subhalo 63875 assembled most of its stellar component at $z > 2$ and was not affected by any mergers since then, the extended star formation episode which started around $z = 1.75$ is not compatible with the expected history of a relic galaxy, because it contaminates this relic candidate with non-negligible young stellar populations.

\subsubsection{Subhalo 282780}
\begin{figure*}
    \centering
    \includegraphics[width=0.9\textwidth]{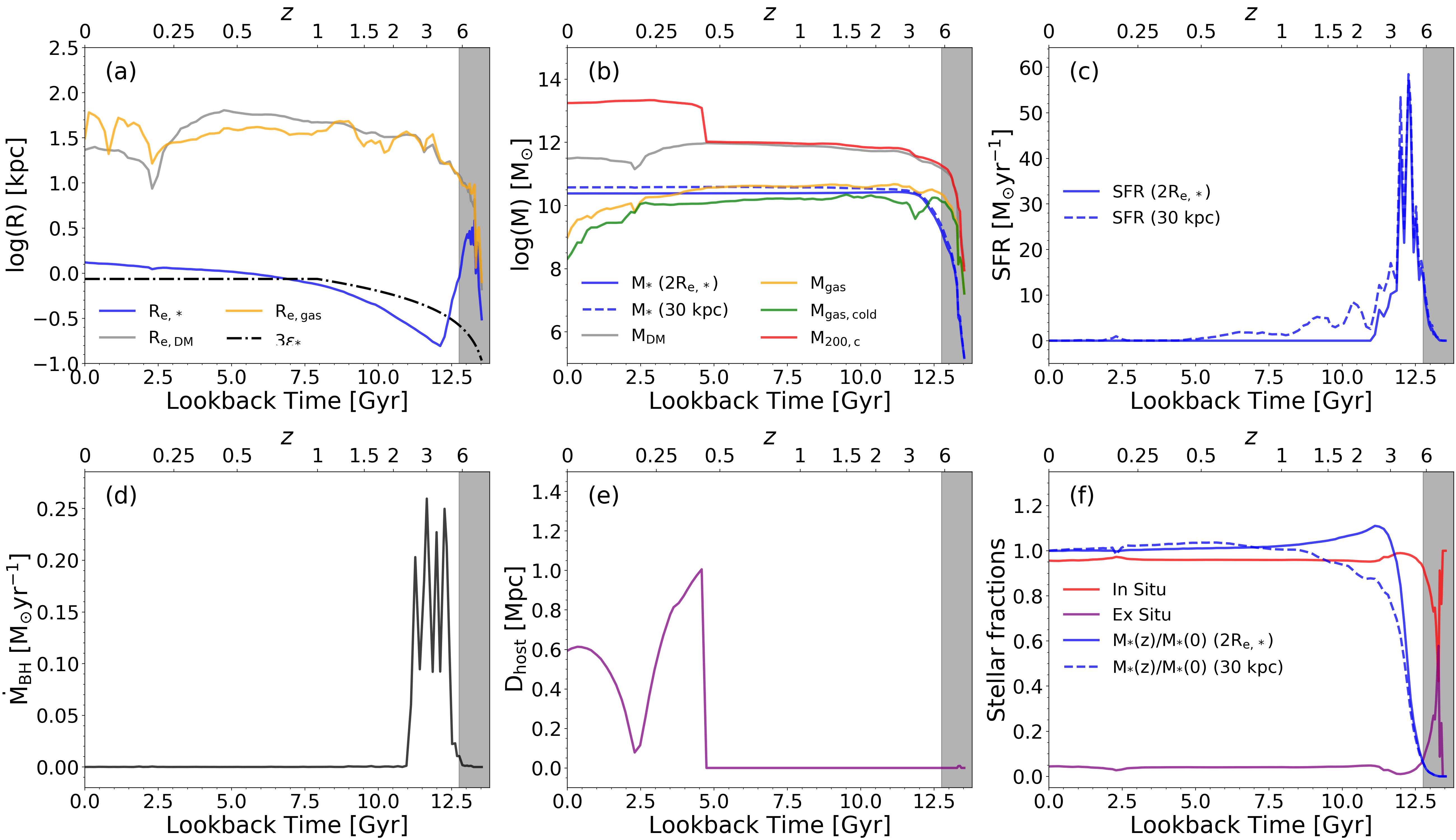}
    \caption{Cosmic history of Subhalo 282780 represented by the evolution of many of its properties. Same as in Fig. \ref{fig:history_6}.}
    \label{fig:history_282780}
\end{figure*}

In Fig. \ref{fig:history_282780}, we present the cosmic evolution for many properties of Subhalo 282780, as in Fig. \ref{fig:history_6}. This candidate formed the bulk of its stellar content very early, reaching already 90 per cent of its $z=0$ stellar mass at $z \sim 2$ (panel f). However, similarly to Subhalo 63875, a prolonged star formation episode in the outer parts ($R > 3$ kpc) consumed part of the surrounding gas until $z \sim 0.5$, giving rise to a contamination of 14 per cent in mass of stellar populations younger than 9.5 Gyr. 

\par This subhalo only recently became a satellite of its current cluster, entering it at $z \sim 0.35$. Similarly to Subhalo 63875, it has a very well-defined gas disk before the first pericentric passage ($z \sim 0.16$), which is subsequently destroyed. During the infall, this simulated galaxy seems to be suffering from ram-pressure stripping, briefly presenting jellyfish-like gas morphology\footnote{Check the videos in Supplementary Material available online.} at $z \sim 0.18$, which is accompanied by a very small and short star formation episode in the outer parts. Although the subhalo made just one pericentric passage, significant size changes are observed in the dark matter component of this simulated galaxy (panel a of Fig. \ref{fig:history_282780}), with $R_{\rm DM}$ decreasing from $\sim$ 60 kpc to $\sim 25$ kpc after the passage. This suggests strong stripping acting upon this subhalo during the close encounter with the central galaxy in the host halo, with the caveat of mass loss ambiguity already mentioned in section \ref{sec:history_sub6}. Nonetheless, this object retained more dark matter and lost relatively less gas compared with other candidates (subhalos 6 and 63875). Regarding the stellar size, we can see in panel (a) that the $R_{\rm e,*}$ increased by about 1~kpc since $z \sim 2$. A small part of this growth can be explained by the star formation in the outer parts of the subhalo which increased its $M_{\rm *}$ by 10 per cent at maximum (see panel c and f). Additionally, it is possible that some effect of artificial expansion due to resolution effect takes place between $z \sim 2$ and $z = 1$, since in this epoch, the gravitational softening length is changing.

\par Comparing the evolution of stellar fractions between the candidates, we see that Subhalo 282780 has the most flat stellar fractions since $z \sim 2$ (panel f), implying little change in its amount of stellar mass, thus favouring the hypothesis of this object being a strong relic analogue. However, as shown Appendix \ref{app:stell_pop}, more than 10 per cent of the mass of this galaxy is made of populations younger than 9.5 Gyr, a non-negligible contamination.

\subsubsection{Subhalo 366407}
\begin{figure*}
    \centering
    \includegraphics[width=0.9\textwidth]{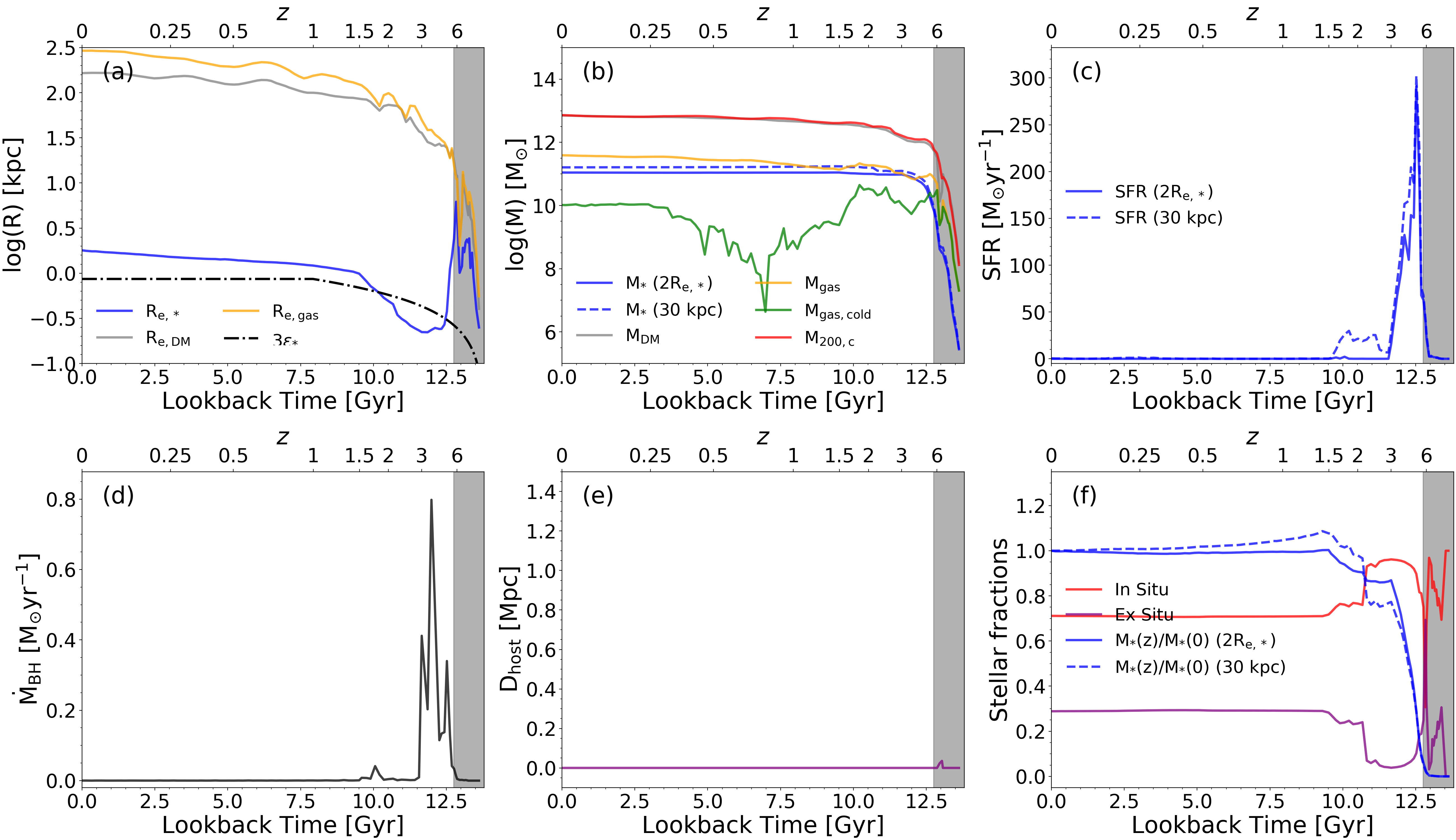}
    \caption{Cosmic history of Subhalo 366407 represented by the evolution of many of its properties. Same as in Fig. \ref{fig:history_6}.}
    \label{fig:history_366407}
\end{figure*}

In Fig. \ref{fig:history_366407}, we present the cosmic evolution for many properties of Subhalo 366407, as in Fig. \ref{fig:history_6}. This subhalo goes through a very early ($z > 3$) star formation burst, as can be seen in panel (c), followed by a weaker but still significant (SFR > 10 $\rm M_\odot$ yr$^{-1}$) star formation episode in its outer parts at $z \sim 2$. Additionally, as shown in panel (d), the first star formation burst at $z > 2$ is accompanied by significant black hole accretion, which then ceases after $z \sim 1.5$. A couple of micro and minor mergers happened between $z=3$ and $z=2$, also increasing the subhalo's stellar mass, which by then reaches 97 per cent of its current value at $z=0$. For the next $\sim 10$ Gyr, this galaxy remains as a central in its parent dark matter halo - check panel (e) - not being affected by mergers and only slightly increasing its $R_{\rm e,*}$ by $\sim 0.6$~kpc, as shown in panel (a). In contrast, since $z \sim 1.5$ this subhalo continuously accrete gas from its surroundings, but does not have any star formation until $z \sim 0.25$, when it experiences a relatively small star formation episode, forming a faint ring of younger stellar particles in the outskirts ($R \sim 30$ kpc). Nonetheless, the amount of stellar mass from populations younger than 9.5 Gyr in this galaxy is only $\sim$1 per cent (see Appendix \ref{app:stell_pop}), which is a very small contamination, not enough to change the overall structure and properties of this relic candidate. Therefore, we consider that the cosmic history of Subhalo 366407 is at least reconcilable with the idealised relic formation scenario. 

\subsubsection{Subhalo 516760}
\begin{figure*}
    \centering
    \includegraphics[width=0.9\textwidth]{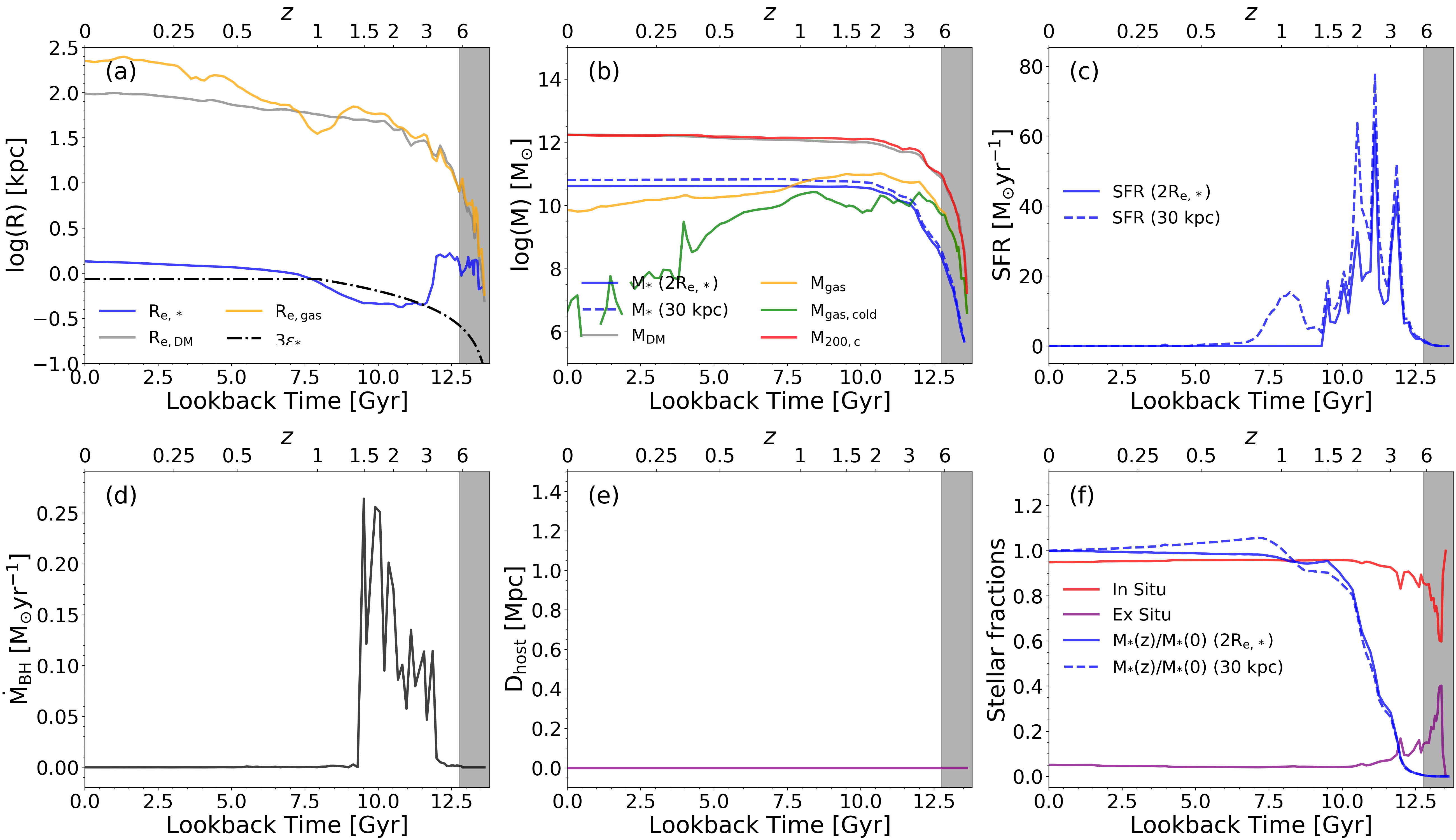}
    \caption{Cosmic history of Subhalo 516760 represented by the evolution of many of its properties. Same as in Fig. \ref{fig:history_6}.}
    \label{fig:history_516760}
\end{figure*}

In Fig. \ref{fig:history_516760}, we present the cosmic evolution for many properties of Subhalo 516760, as in Fig. \ref{fig:history_6}. This galaxy remains a central galaxy in its halo and evolves mostly isolated during its history, only having a micro merger at $z \sim 0.35$. It has a late star formation episode in the outskirts at $z \sim 1$, which increased the total $M_{\rm *}$ by $\sim$13 per cent, but part of this increase was gradually reverted until $z \sim 0$.
Regarding the gas component, we can see in panel (a) of Fig. \ref{fig:history_516760} that the half-mass radius of gas continuously increases since $z \sim 0.8$. However, this should not be taken as a radius increase of a symmetric gas structure during the time interval involved. After the micro merger with a small satellite at $z \sim 0.36$, the subhalo forms a kind of gas tail (check videos in Supplementary Material), which results in a very large size for the gas component. Any increase in gas mass possibly associated with an increase in gas radius is ruled out if we look at the panel (b), where we see that the gas mass decreases since $z \sim 0.8$. This object's relatively late ($z \sim 1$) star formation peak result in a mass contribution of young stellar populations that goes over 20 per cent (see Appendix \ref{app:stell_pop}). Considering also the recent micro merger at $z \sim 0.36$, this candidate must inevitably be disregarded as a strong relic analogue.

\section{Discussion}
In Subsection \ref{sec:sample_selection}, we select 48 relic analogue candidates from TNG50 and after analysing the stellar assembly, size evolution and stellar surface density, we narrow down to 5 subhalos that could be potential relics. Here, we summarise the results presented in section \ref{sec:Results}, judging which of the 5 subhalos candidates could represent strong relic analogues in TNG50. More specifically, we consider strong relic analogues the candidates which have properties compatible to what is expected from observations and that did not have significant changes in their stellar population or structure since their formation ($z > 1.5$). We refer here to the degree of relicness, which is a measure of how much a relic galaxy has preserved its red nugget properties. In section \ref{sec:relicness}, we describe the score adopted to quantify the relicness of the 5 simulated relic candidates analysed throughout this work and, in Table \ref{tab:relicness}, we present the relicness score for each candidate.

\subsection{Which are the best relic analogues in TNG50?}\label{sec:best_relics}
\par The 5 simulated relic analogue candidate galaxies analysed in the previous section do not suffer any major or minor mergers since $z \sim 1.7$ and also have super-solar stellar metallicities and [Mg/Fe]. This is in agreement with the proposed scenario in which a relic galaxy would form rapidly at high redshift and then evolve passively until $z=0$. 

\begin{table*}
\centering
\caption{Criteria used to classify the relic analogue candidates considering a degree of relicness, for details on the definition of the relicness score, see section \ref{sec:relicness}. \textit{Weak age uniformity}: $\gtrsim$ 75 per cent of $M_*$ composed of stellar populations older than 9.5 Gyr. \textit{Super-solar $Z$ \& [Mg/Fe]}: super-solar average metallicity and [Mg/Fe]. \textit{Compact \& high $\sigma_{\rm c}$}: $R_{\rm e,*}$ below median at similar $M_*$ and $\sigma_{\rm c} >$ 200~km~s\textsuperscript{-1}. \textit{Strong age uniformity}: $\geq$ 95 per cent of $M_*$ composed of stellar populations older than 9.5 Gyr. \textit{No young contaminant}: no stellar populations younger than 3 Gyr. Subhalo 6 is the only candidate which satisfy all the criteria, being the strongest relic galaxy analogue among the candidates.} 
\begin{tabular}{ccccccc} \hline
ID & Weak age uniformity & Super-solar $Z$ \& [Mg/Fe] & Compact \& high $\sigma_{\rm c}$ & Strong age uniformity & No young population & Relicness score \\
6 & Y & Y & Y & Y & Y & 5 \\
63875 & Y & Y & N & N & Y & 2 \\
282780 & Y & Y & N & N & N & 2 \\
366407 & Y & Y & Y & Y & N & 4 \\
516760 & Y & Y & N & N & Y & 2 \\ 
\hline
\end{tabular}
\label{tab:relicness}
\end{table*}

\par Subhalos 63875 and 282780 are satellite galaxies and also the less massive candidates (see Table \ref{tab:properties}), having overall properties closer to the median values of other quiescent galaxies with similar $M_{\rm *}$ (check scatter plots of section \ref{sec:global_properties}). Even though Subhalo 282780 formed very early and is the second oldest in the candidates sample, it is the one with the lowest $M_{\rm *}$ and $\sigma_{\rm c}$. Relaxing the constraints imposed by Y17 sample, towards less extreme $\sigma_{\rm c}$ and $M_{\rm *}/M_{\rm dyn}$, Subhalo 282780 could maybe be considered an intermediate-mass relic analogue, as the one discovered by \cite{FerreMateu2018}, PGC012519. However, Subhalo 282780 has non-negligible star formation activity extending until $z \sim 1$, so it would have a lower degree of relicness. Moreover, $>$14 per cent of its stellar mass is made from stellar populations younger than 9.5 Gyr, a considerable mass fraction of stars that are not old, another property disfavouring this object as a strong relic analogue. Subhalo 63875 presents an even more extended and significant star formation activity (panel (c) in Fig. \ref{fig:history_63875}), with an even higher mass fraction of young stellar populations. Additionally, this candidate also exhibits a peculiar pattern in its isophotes and velocity maps, having what seems to be a disk that is warped or distorted (check Fig \ref{fig:H-band} and Fig. \ref{fig:kinematics}). 
The similarity of subhalos 282780 and 63875 with real relics is more due to their early stellar assembly than to their global properties and cosmic histories, as the relicness score suggests. Thus, we do not consider these objects as strong relic analogues.
\par The youngest candidate in the sample, Subhalo 516760, is a central galaxy in its local environment and only accreted residual amounts of stellar mass after $z \sim 1.5$. It deviates in many of the plots presented in subsection \ref{sec:global_properties}, in a direction which is in agreement with observational expectations for relics, e.g. being very compact, metal-rich and strongly dominated by stars at $R<1 R_{\rm e,*}$ (see Fig. \ref{fig:properties}). However, it has also been recently ($z\sim0.36$) crossed by a very small satellite, which affects its inner structure, thus, decreasing its degree of relicness. In addition, this subhalo has the most extended star formation activity among the candidates, being the most proportionally contaminated by young stellar populations, in terms of mass (check Appendix \ref{app:stell_pop}). The morphology of this subhalo, as shown in Fig. \ref{fig:H-band} is also visually discrepant from the observed relic galaxies, presenting some boxy isophotes as opposed to disky, like the galaxies from Y17 or the red nuggets observed at high-$z$ \citep{vanderWel2011}. This simulated galaxy has an evident rotation disk, but its central velocity dispersion is slightly lower ($\sigma_{\rm c} \sim $~175~km~s\textsuperscript{-1}, see Fig. \ref{fig:kinematics}) than what is expected from the Y17 sample. The star formation activity that happened after $z \sim 1.5$ clearly left an impact in the age profile of this candidate, decreasing its uniformity in age and consequently decreasing its degree of relicness. We consider that this candidate has enough properties that differ from the observational and theoretical expectations for relic galaxies, which is captured by the relicness score (see Table \ref{tab:relicness}) defined in section \ref{sec:relicness}. Thus, we disregard Subhalo 516760 as a strong relic analogue - even in the context of a degree of relicness. In contrast, Subhalo 366407 is a more promising candidate. Looking at its global properties, we already noticed that it follows almost all the observational expectations of relic galaxies, being very compact, metal-rich, alpha-enhanced and strongly dominated by stars at $R<1 R_{\rm e,*}$ (see Fig. \ref{fig:properties}). This candidate is a central galaxy in its parent dark matter halo and evolves passively until $z \sim 0.25$, when it presents some negligible star formation activity in its outskirts ($R \sim 30$~kpc), slightly decreasing its degree of relicness. However, the mass fraction of stellar particles younger than 9.5 Gyr in this subhalo is only $\sim$1\%, the lowest among the candidates (check Appendix \ref{app:stell_pop}). Apart from the very faint young stellar ring formed in the outskirts, this subhalo is overall morphologically similar to real relics, having disky isophotes as shown in Fig. \ref{fig:H-band}. Additionally, from the kinematic maps in Fig. \ref{fig:kinematics}, we see that there is a clear well-defined rotation pattern in this object and also that its central $\sigma$ reaches at least 300~km~s\textsuperscript{-1}, both features in agreement with characteristics of real relics from Y17. Subhalo 366407 is also the oldest object among the candidates, with a very flat age profile, similar to Subhalo 6 (see  panel (c) in Fig. \ref{fig:profiles}). Given all these properties, we consider that Subhalo 366407 represents a strong relic analogue with an intermediate degree of relicness, since it already started its path to become more similar to an ETG, analogous to what is suggested by \cite{FerreMateu17} in the case of MRK 1216.
\par Finally, Subhalo 6, which is a satellite galaxy in a cluster environment, seems to be the best relic analogue candidate in our sample. It is a uniformly old object, has super-solar metallicity and [Mg/Fe] out to 4$R_{\rm e,*}$, disky isophotes, a minimum $\sigma_{\rm c}$ of 200 km/s, a clear rotation disk, and high $M_{*}/M_{\rm dyn}$ within $1R_{\rm e,*}$ ($\sim 0.84$); all properties in agreement to what is expected from the observations of real relics. Additionally, it is close to the centre of the most massive cluster in the simulation and has a very low dark matter-to-stellar mass fraction (log $M_{\rm DM}/M_{\rm *} = 0.15$). This is peculiarly similar to NGC 1277, that also has low dark matter-to-stellar mass fraction (log $M_{\rm DM}/M_{\rm *} = -0.91^{+3.58}_{-1.44}$) according to orbit-based dynamical models from \cite{Yildirim2015,yildirim17} and which is a confirmed relic residing close to the centre of the Perseus Cluster (Abell 426) - one of the closest massive clusters known. Although the constraints on the dark matter fraction are poor - as one can see by the uncertainties (see Table \ref{tab:obs_properties}) - the similarity in environment to NGC 1277 is interesting. Since it satisfies most of the observational constraints imposed by Y17 sample, and, its stellar structure and populations have not changed significantly since $z \sim 1.5$, we consider that Subhalo 6 represents a strong relic analogue.

\subsection{Number density}
Other authors have shown that relic galaxies are more likely to be found in denser environments \citep{Poggianti2013b,Stringer2015,Peralta2016}, so the presence of a relic analogue in a massive cluster of the simulation would not be surprising if the small simulated Universe of TNG50 reproduced well the number densities of different classes of galaxies. Previous studies predict relic number densities at $z \sim 0$ that are of the order of $\sim 6\times 10^{-7}$ Mpc$^{-3}$ \citep{FerreMateu17,Quilis2013}, if we apply this estimate to the volume of TNG50, we would expect to find less than one relic in the simulation ($\sim 0.1$). However, the real relic number density may be a function environmental density, likely being higher in clusters. Since TNG50 is already able to produce massive ($M > 10^{14} \ \rm M_{\odot}$) clusters, this can possibly explain why we could find at least one relic-like object in the simulation (Subhalo 6), despite its relatively small comoving volume. If we consider that the 13 observed galaxies of Y17 used in this work are good relic representatives, and compute the relic number density inside a sphere containing all of them (with radius $R$ = 113 $\pm$ 2), we estimate a value of $\sim 2 \times 10^{-6}$ Mpc$^{-3}$. Considering that we find at least 1 strong relic analogue and that the volume of TNG50 box at z = 0 is 51.7 Mpc$^3$, a crude estimate of the relic number density in the simulation is $7 \times 10^{-6}$. This is still larger than the observational estimate, but less than an order of magnitude different. Furthermore, the observational estimates could be affected mainly by two issues, first, a bias due to lack of completeness in the sample adopted as reference, and second, the distance estimates assumed here. Additionally, it could also be that TNG50-1 simulation overestimates the relic number density, perhaps due to a more efficient early quenching that may not happen in the population of real relic galaxies at $z > 1.5$. The ongoing INSPIRE survey may provide a more complete sample of relic galaxies with reliable distance estimates, constraining relic number density at different redshifts; consequently, the results of this work can possibly be tested in the near future. 

\subsection{Formation of the relic analogues}
 Previous work by \cite{wellons15}, argue that compact elliptical galaxies are just extreme representatives of a smooth distribution of galaxy properties, being an expected outcome from the different combinations of physical processes galaxies may undergo. Additionally, in Illustris, \cite{wellons16} found massive compact galaxies at $z \sim 0$ that remained nearly undisturbed since $z \sim 2$, and predict that these compact relics should predominantly live in under-dense environments or as satellites in larger groups at the local Universe. The formation of the best relic analogue - Subhalo 6 - found in this work also emphasises the role of environment, since almost all the remaining gas of Subhalo 6 was removed at early epochs during its infall into its parent halo. The combination of this gas supply removal and the very early substantial star formation, naturally gives rise to the relic nature of this object in the simulation and could possibly be one formation pathway to explain the formation of the observed real relics. In fact, \cite{FerreMateu17} and other authors already suggested that the relics with most extreme properties are expected to be in rich environments, such as the cluster in which Subhalo 6 resides. Our finding thus provide a direct support for this hypothesis drawn from an observational perspective. The second best relic analogue found in TNG50, Subhalo 366407, has age, metallicity and [Mg/Fe] profiles similar to Subhalo 6, but goes through very different environmental interactions throughout its history. First, Subhalo 366407 is a central galaxy in its local environment, thus, not suffering stripping of its gas. Instead, the halt in its star formation activity arises from the depletion of cold gas. Possibly as a consequence of this different origin without gas stripping, the density profile of this relic analogue is more extended than the one of Subhalo 6 (see panel (d) in Fig. \ref{fig:profiles}). 
 \par From an observational perspective, a broader explanation on the diverse formation pathways of compact elliptical galaxies has been presented by \cite{FerreMateu2018} and \cite{Ferre-Mateu2021}. Here we present results indicating that the relic analogues in TNG50 - a subset of compact quiescent galaxies in the simulation - can also be formed by distinct pathways depending on the environment, in alignment with the hypothesis proposed by \cite{FerreMateu17}.

\subsection{Notes on primordial halo properties}
One important aspect, but still difficult to explore directly in observations, is the formation environment of relic galaxies and other compact quiescent galaxies in general. Fortunately, this aspect can be partially explored through the use of cosmological simulations. In this section, we briefly discuss some simpler properties of the relic analogue candidates primordial host halos, such as virial mass and gas mass fraction. 

\begin{figure}
    \centering
    \includegraphics[width=0.47\textwidth]{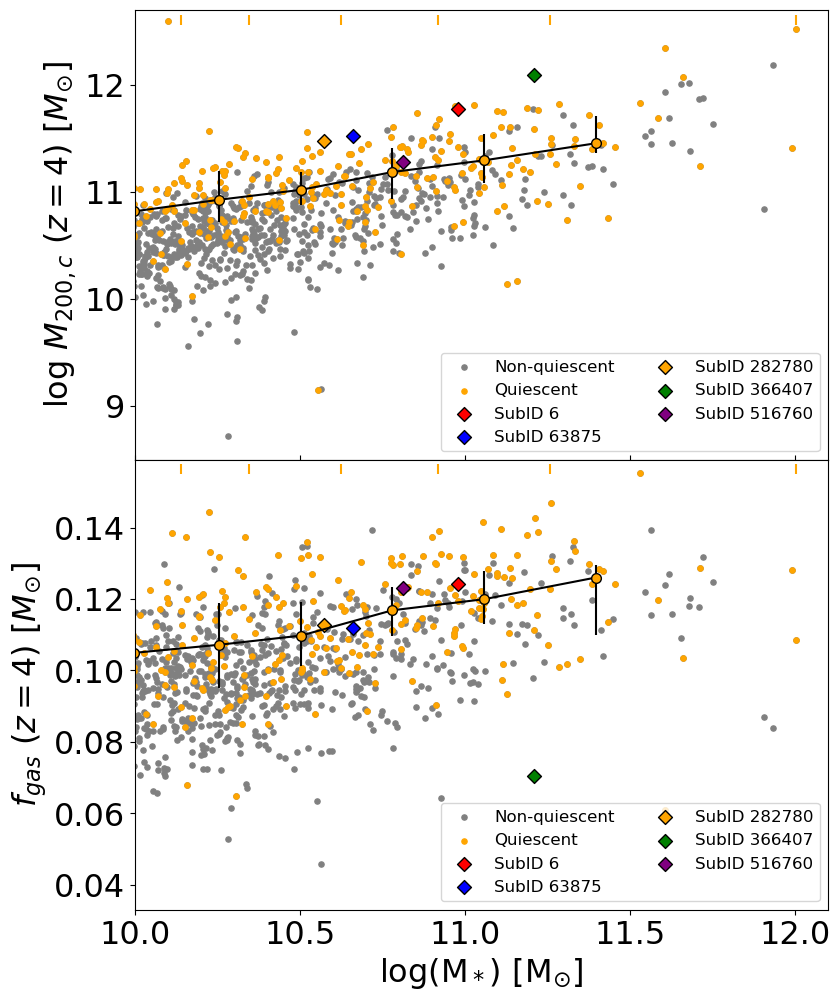}
    \caption{Properties of the parent halos at $z=4$. $M_{\rm *}$ corresponds to the stellar mass of the subhalo at $z=0$. Top panel shows the virial mass ($M_{\rm 200,c}$) and in the bottom panel, the total gas mass fraction ($f_{\rm gas}$). Coloured diamonds represent relic analogue candidates, orange dots are quiescent objects, grey points are non-quiescent objects. Median values for quiescent galaxies are computed for each mass bin and represented by the larger orange circles connected by black solid lines, error bars are simply the 25th and 75th percentiles for each mass bin. The stellar mass bins edges are represented by the orange ticks at the top of the panels, as in Fig. \ref{fig:properties}}
    \label{fig:scat_env}
\end{figure}

In Fig. \ref{fig:scat_env}, we show the properties of the parent halos which hosted the $z=4$ progenitors of the relic candidates found at $z=0$. In the top panel of this figure, we can see that the relic analogues - Subhalo 6 and Subhalo 366407 - at $z=4$ seem to be in relatively more massive halos than other quiescent galaxies of similar mass at $z=0$. Regarding the gas mass fraction in the parent halos, in the bottom panel of the figure, we see that Subhalo 6 does not deviate strongly from the median values of quiescent population. In contrast, Subhalo 366407 is far below the median values of other quiescent subhalos with similar mass at $z=0$. Considering that this candidate is the oldest relic candidate in the sample, it may have already consumed significant amounts of its surrounding gas at $z = 4$, thus explaining its behaviour in the lower panel of Fig. \ref{fig:scat_env}.
Our results regarding the virial mass, are aligned with the idea of old galaxies observed at $z \sim 0$ being formed in the most massive dark matter halos at high-redshift. In this sense, the relic galaxies found in the nearby Universe would be just the extreme case of old galaxies with a head start, that formed first and were not significantly contaminated by younger stellar populations afterwards. 
\par A recent work using the TNG50 simulation \citep{Du2021} speculates that the initial spin of parent dark matter halos may play an important role, controlling the size evolution of galaxies. The primordial spin thus could have an impact in the formation of relic analogues as those found here, however, this kinematic analysis is considerably more complex, lying beyond the scope of this work.

\section{Summary and Conclusions}\label{sec:Conclusions}
ETGs are the most massive galaxies observed today and the exact mechanisms responsible for their formation are still subject of great debate. In the simplified two-phase formation scenario, ETGs should experience a very intense stage in terms of stellar assembly at high-redshift ($z > 2$), with their progenitors being very compact, massive and quiescent already at $z \sim 2$. If a galaxy undergoes only through this first phase and survives unaltered until today ($z=0$), it is called a relic galaxy. These rare objects are very useful for the study of galaxy evolution, because they allow us to have a glimpse in the formation of ETGs, acting like a window to the high-redshift Universe. 
\par In order to understand what influences the formation of a relic galaxy, we searched for relic galaxy analogues in one of the cosmological simulations from the IllustrisTNG Project, pushing the limits of current state-of-the-art cosmological simulations in terms of compact galaxy formation. We consider as strong relic analogues, the objects at $z=0$ in the simulation that have global properties similar to objects in the observational sample used as reference in this work, in addition to not having their stellar components significantly altered since $z \sim 1.5$. 
\par We use the highest resolution simulation run available in the IllustrisTNG Project (TNG50-1) and start by searching for old compact massive quiescent galaxies at $z=0$. After applying different criteria for compactness, quiescence and age, we find 48 relic analogue candidates. Once confirmed that TNG50 could generate such compact simulated galaxies, we filter this first sample, selecting a final sample of objects that had more than 75 per cent of their mass composed of populations older than 9.5 Gyr and that did not suffer significant stellar size changes since $z \sim 1.5$. Only 5 subhalos remain and are considered in this work as the best relic analogue candidates. Then, we check if any of these objects could be considered relic galaxy analogues by comparing their properties with what is expected from observational works - like \cite{yildirim17} - and the idealised scenario for the formation of relics. We also analyse the formation histories of the relic analogue candidates, in order to determine the main physical mechanisms which prevented them to significantly increase stellar masses and sizes over a period of more than 9 Gyr. 
\par We find that the TNG50 simulation is already able to produce at least one simulated galaxy with many properties similar to real relic galaxies at $z \sim 0$. Thus, we consider that this object - Subhalo 6 - represents the strongest relic analogue. This is motivated by the fact that Subhalo 6 is the candidate which satisfies most of the constraints imposed by the observational sample used as reference in this work, and even has a peculiar similarity with one of the real relic galaxies (NGC 1277). Another subhalo found in this work - Subhalo 366407 - can be considered as a strong relic analogue if a certain degree of relicness is accepted. Subhalo 366407 has stellar mass, stellar size, metallicity, morphology and central velocity dispersion in accordance with being a relic analogue. Although Subhalo 366407 has a stellar assembly and star formation very similar to what is expected from the idealised scenario of a relic, it has a small contamination of young (age < 3 Gyr) stellar populations in its outskirts ($\sim 1 \%$ in mass), originated from recent gas accretion. This contamination is the reason we invoke the degree of relicness for Subhalo 366407, since it already started to assemble relatively young stellar populations. Another 3 simulated galaxies in TNG50 were studied as potential relic galaxy analogues, but because they do not satisfy most observational constraints adopted in this work (see section \ref{sec:best_relics}), we consider that they may not meaningfully represent the evolution of real relic galaxies. 
\par Nevertheless, none of the relic analogue candidates simultaneously satisfy \textit{all} the constraints imposed by \cite{yildirim17} sample. All relic candidates in TNG50 have central regions less dense than observations, which we suspect may be caused by numerical resolution issues, as was already suggested by recent work \citep{Zanisi2021}. Therefore, we stress the fact that the objects found in the simulation are not \textit{perfect} analogues, since their internal structure may not fully reflect the ones of real galaxies. Also, relic galaxies are rare in the real Universe, thus, it would not be surprising to not find any of them in the small comoving box simulated in TNG50. However, the finding of at least one strong relic analogue in TNG50 is in rough agreement with our observational number density estimation (2$\times 10^{-6} \rm Mpc^{-3}$), which considers the 13 galaxies in the observational sample adopted as reference here.
\par Regarding the formation of relics in TNG50, we conclude that the two relic analogues described here represent two distinct formation pathways for relic galaxies. These are closely connected to the environment in which the relic progenitors evolved, as previously discussed in observational works \citep{FerreMateu17}. Subhalo 6 might represent real relic galaxies which are satellites in dense environments, while Subhalo 366407 might represent the relics which evolved more isolated. We also briefly explore properties of the halos in which these relic candidates formed, checking if their progenitors formed in gas-rich and massive host halos. We find that the relic analogues seem to be formed in more massive halos than other quiescent galaxies with similar stellar mass at $z=0$. Regarding the gas fraction, we find that relic analogues do not follow any clear trend with respect to other simulated quiescent galaxies of similar stellar mass at $z=0$. It is important to note that TNG was tuned to match $z=0$ observations, so a better match is expected at this redshift. However, any conclusions we draw for high-redshift ($z > 1$) are directly affected by the cosmological and sub-grid models adopted in the simulation.
\par In summary, we study one of the most recent cosmological simulations and compare its results to observations, aiming to understand the formation of relic galaxies and test the ability of the simulations to reproduce such rare objects. Future high-redshift observations and relic-directed spectroscopic surveys will help to constrain even more the first stages of ETG assembly. The next generation of cosmological simulations with novel techniques, higher resolution and larger volumes will provide new insights into the open questions of galaxy formation and evolution. This work aims to connect observations with simulations, an approach that we believe can be very fruitful, since it promotes a synergy between these two successful branches of modern astronomy and astrophysics.  

\section*{Acknowledgements}
The authors thank the anonymous referee for their comments and suggestions which led to an improved version of the manuscript. This work is based on an undergraduate thesis as part of the requirements for obtaining the title of Bachelor's in Physics: Astrophysics at \textit{Universidade Federal do Rio Grande do Sul} and written under the COVID-19 pandemic remote work period. The authors acknowledge the helpful and insightful discussions with Michael Beasley and Rubens Machado. RFF acknowledge the technical support of the TNG Team, specially Dylan Nelson. ACS acknowledge funding from the brazilian agencies \textit {Conselho Nacional de Desenvolvimento Cient\'ifico e Tecnol\'ogico} (CNPq) and the  \textit{Fundação de Amparo à Pesquisa do Estado do RS} (FAPERGS) through grants CNPq-403580/2016-1, CNPq-11153/2018-6, PqG/FAPERGS-17/2551-0001, FAPERGS/CAPES 19/2551-0000696-9, L'Or\'eal UNESCO ABC \emph{Para Mulheres na Ci\^encia} and the Chinese Academy of Sciences (CAS) President's International Fellowship Initiative (PIFI) through grant E085201009. CF acknowledges the financial support from CNPq through grants CNPq-433615/2018-4 e CNPq-314672/2020-6. The IllustrisTNG simulations were undertaken with compute time awarded by the Gauss Centre for Supercomputing (GCS) under GCS Large-Scale Projects GCS-ILLU and GCS-DWAR on the GCS share of the supercomputer Hazel Hen at the High Performance Computing Center Stuttgart (HLRS), as well as on the machines of the Max Planck Computing and Data Facility (MPCDF) in Garching, Germany. This work was also based on observations made with the NASA/ESA Hubble Space Telescope, and obtained from the Hubble Legacy Archive, which is a collaboration between the Space Telescope Science Institute (STScI/NASA), the Space Telescope European Coordinating Facility (ST-ECF/ESA) and the Canadian Astronomy Data Centre (CADC/NRC/CSA). 

\section*{Data Availability}
The IllustrisTNG simulations, including TNG50-1, are publicly available at \url{www.tng-project.org/data} \citep{Nelson2019b}. Images from HST used in this publication are publicly available in the Hubble Legacy Archive.
Additional data directly related to this publication are available on request from the corresponding author.

\section*{Code Availability}
 In this work, we used Python libraries: \textsc{Astropy}, \textsc{NumPy}, \textsc{Numba}, \textsc{Pandas} and \textsc{SciPy}. We also used the \textsc{illustris\_python} module, available at \url{www.github.com/illustristng/illustris_python}.
 Additional code used to produce this publication are available on reasonable request from the corresponding author. 



\bibliographystyle{mnras}
\bibliography{paper} 

\begin{thebibliography}{}
\makeatletter
\relax
\def\mn@urlcharsother{\let\do\@makeother \do\$\do\&\do\#\do\^\do\_\do\%\do\~}
\def\mn@doi{\begingroup\mn@urlcharsother \@ifnextchar [ {\mn@doi@}
  {\mn@doi@[]}}
\def\mn@doi@[#1]#2{\def\@tempa{#1}\ifx\@tempa\@empty \href
  {http://dx.doi.org/#2} {doi:#2}\else \href {http://dx.doi.org/#2} {#1}\fi
  \endgroup}
\def\mn@eprint#1#2{\mn@eprint@#1:#2::\@nil}
\def\mn@eprint@arXiv#1{\href {http://arxiv.org/abs/#1} {{\tt arXiv:#1}}}
\def\mn@eprint@dblp#1{\href {http://dblp.uni-trier.de/rec/bibtex/#1.xml}
  {dblp:#1}}
\def\mn@eprint@#1:#2:#3:#4\@nil{\def\@tempa {#1}\def\@tempb {#2}\def\@tempc
  {#3}\ifx \@tempc \@empty \let \@tempc \@tempb \let \@tempb \@tempa \fi \ifx
  \@tempb \@empty \def\@tempb {arXiv}\fi \@ifundefined
  {mn@eprint@\@tempb}{\@tempb:\@tempc}{\expandafter \expandafter \csname
  mn@eprint@\@tempb\endcsname \expandafter{\@tempc}}}

\bibitem[\protect\citeauthoryear{{Alamo-Mart{\'\i}nez}
  et~al.,}{{Alamo-Mart{\'\i}nez} et~al.}{2021}]{Alamo-Martinez2021}
{Alamo-Mart{\'\i}nez} K.~A.,  et~al., 2021, \mn@doi [\mnras]
  {10.1093/mnras/stab538}, \href
  {https://ui.adsabs.harvard.edu/abs/2021MNRAS.503.2406A} {503, 2406}

\bibitem[\protect\citeauthoryear{{Asplund}, {Grevesse}, {Sauval}  \&
  {Scott}}{{Asplund} et~al.}{2009}]{asplund2009}
{Asplund} M.,  {Grevesse} N.,  {Sauval} A.~J.,   {Scott} P.,  2009, \mn@doi
  [\araa] {10.1146/annurev.astro.46.060407.145222}, \href
  {https://ui.adsabs.harvard.edu/abs/2009ARA&A..47..481A} {47, 481}

\bibitem[\protect\citeauthoryear{{Barbosa}, {Spiniello}, {Arnaboldi},
  {Coccato}, {Hilker}  \& {Richtler}}{{Barbosa} et~al.}{2021}]{Barbosa2021}
{Barbosa} C.~E.,  {Spiniello} C.,  {Arnaboldi} M.,  {Coccato} L.,  {Hilker} M.,
    {Richtler} T.,  2021, \mn@doi [\aap] {10.1051/0004-6361/202039809}, \href
  {https://ui.adsabs.harvard.edu/abs/2021A&A...649A..93B} {649, A93}

\bibitem[\protect\citeauthoryear{{Barro} et~al.,}{{Barro}
  et~al.}{2013}]{Barro2013}
{Barro} G.,  et~al., 2013, \mn@doi [\apj] {10.1088/0004-637X/765/2/104}, \href
  {https://ui.adsabs.harvard.edu/abs/2013ApJ...765..104B} {765, 104}

\bibitem[\protect\citeauthoryear{{Beasley}, {Trujillo}, {Leaman}  \&
  {Montes}}{{Beasley} et~al.}{2018}]{Beasley2018}
{Beasley} M.~A.,  {Trujillo} I.,  {Leaman} R.,   {Montes} M.,  2018, \mn@doi
  [\nat] {10.1038/nature25756}, \href
  {https://ui.adsabs.harvard.edu/abs/2018Natur.555..483B} {555, 483}

\bibitem[\protect\citeauthoryear{{Bertin} \& {Arnouts}}{{Bertin} \&
  {Arnouts}}{1996}]{Bertin96}
{Bertin} E.,  {Arnouts} S.,  1996, \mn@doi [\aaps] {10.1051/aas:1996164}, \href
  {https://ui.adsabs.harvard.edu/abs/1996A&AS..117..393B} {117, 393}

\bibitem[\protect\citeauthoryear{{Bezanson}, {van Dokkum}, {Tal}, {Marchesini},
  {Kriek}, {Franx}  \& {Coppi}}{{Bezanson} et~al.}{2009}]{Bezanson09}
{Bezanson} R.,  {van Dokkum} P.~G.,  {Tal} T.,  {Marchesini} D.,  {Kriek} M.,
  {Franx} M.,   {Coppi} P.,  2009, \mn@doi [\apj]
  {10.1088/0004-637X/697/2/1290}, \href
  {https://ui.adsabs.harvard.edu/abs/2009ApJ...697.1290B} {697, 1290}

\bibitem[\protect\citeauthoryear{{Bournaud}, {Jog}  \& {Combes}}{{Bournaud}
  et~al.}{2007}]{Bournaud2007}
{Bournaud} F.,  {Jog} C.~J.,   {Combes} F.,  2007, \mn@doi [\aap]
  {10.1051/0004-6361:20078010}, \href
  {https://ui.adsabs.harvard.edu/abs/2007A&A...476.1179B} {476, 1179}

\bibitem[\protect\citeauthoryear{Bradley et~al.,}{Bradley
  et~al.}{2020}]{bradley2020}
Bradley L.,  et~al., 2020, astropy/photutils: 1.0.0,
  \mn@doi{10.5281/zenodo.4044744}, \url
  {https://doi.org/10.5281/zenodo.4044744}

\bibitem[\protect\citeauthoryear{{Bruzual} \& {Charlot}}{{Bruzual} \&
  {Charlot}}{2003}]{GALEXEV}
{Bruzual} G.,  {Charlot} S.,  2003, \mn@doi [\mnras]
  {10.1046/j.1365-8711.2003.06897.x}, \href
  {https://ui.adsabs.harvard.edu/abs/2003MNRAS.344.1000B} {344, 1000}

\bibitem[\protect\citeauthoryear{{Buitrago}, {Trujillo}, {Conselice},
  {Bouwens}, {Dickinson}  \& {Yan}}{{Buitrago} et~al.}{2008}]{Buitrago2008}
{Buitrago} F.,  {Trujillo} I.,  {Conselice} C.~J.,  {Bouwens} R.~J.,
  {Dickinson} M.,   {Yan} H.,  2008, \mn@doi [\apjl] {10.1086/592836}, \href
  {https://ui.adsabs.harvard.edu/abs/2008ApJ...687L..61B} {687, L61}

\bibitem[\protect\citeauthoryear{{Buitrago} et~al.,}{{Buitrago}
  et~al.}{2018}]{buitrago2018}
{Buitrago} F.,  et~al., 2018, \mn@doi [\aap] {10.1051/0004-6361/201833785},
  \href {https://ui.adsabs.harvard.edu/abs/2018A&A...619A.137B} {619, A137}

\bibitem[\protect\citeauthoryear{{Buote} \& {Barth}}{{Buote} \&
  {Barth}}{2019}]{Buote2019}
{Buote} D.~A.,  {Barth} A.~J.,  2019, \mn@doi [\apj]
  {10.3847/1538-4357/ab1008}, \href
  {https://ui.adsabs.harvard.edu/abs/2019ApJ...877...91B} {877, 91}

\bibitem[\protect\citeauthoryear{Camps \& Baes}{Camps \& Baes}{2015}]{SKIRT8}
Camps P.,  Baes M.,  2015, \mn@doi [Astronomy and Computing]
  {10.1016/j.ascom.2014.10.004}, 9, 20

\bibitem[\protect\citeauthoryear{Camps \& Baes}{Camps \&
  Baes}{2020}]{Camps2020SKIRTGrains}
Camps P.,  Baes M.,  2020, \mn@doi [Astronomy and Computing]
  {10.1016/j.ascom.2020.100381}, 31

\bibitem[\protect\citeauthoryear{{Cappellari}}{{Cappellari}}{2016}]{Cappellari2016}
{Cappellari} M.,  2016, \mn@doi [\araa] {10.1146/annurev-astro-082214-122432},
  \href {https://ui.adsabs.harvard.edu/abs/2016ARA&A..54..597C} {54, 597}

\bibitem[\protect\citeauthoryear{{Cappellari} et~al.,}{{Cappellari}
  et~al.}{2011}]{Capellari2011}
{Cappellari} M.,  et~al., 2011, \mn@doi [\mnras]
  {10.1111/j.1365-2966.2010.18174.x}, \href
  {https://ui.adsabs.harvard.edu/abs/2011MNRAS.413..813C} {413, 813}

\bibitem[\protect\citeauthoryear{{Daddi} et~al.,}{{Daddi}
  et~al.}{2005}]{Daddi2005}
{Daddi} E.,  et~al., 2005, \mn@doi [\apj] {10.1086/430104}, \href
  {https://ui.adsabs.harvard.edu/abs/2005ApJ...626..680D} {626, 680}

\bibitem[\protect\citeauthoryear{{Damjanov} et~al.,}{{Damjanov}
  et~al.}{2009}]{Damjanov2009}
{Damjanov} I.,  et~al., 2009, \mn@doi [\apj] {10.1088/0004-637X/695/1/101},
  \href {https://ui.adsabs.harvard.edu/abs/2009ApJ...695..101D} {695, 101}

\bibitem[\protect\citeauthoryear{{Dav{\'e}}, {Angl{\'e}s-Alc{\'a}zar},
  {Narayanan}, {Li}, {Rafieferantsoa}  \& {Appleby}}{{Dav{\'e}}
  et~al.}{2019}]{Dave2019}
{Dav{\'e}} R.,  {Angl{\'e}s-Alc{\'a}zar} D.,  {Narayanan} D.,  {Li} Q.,
  {Rafieferantsoa} M.~H.,   {Appleby} S.,  2019, \mn@doi [\mnras]
  {10.1093/mnras/stz937}, \href
  {https://ui.adsabs.harvard.edu/abs/2019MNRAS.486.2827D} {486, 2827}

\bibitem[\protect\citeauthoryear{{Dekel} et~al.,}{{Dekel}
  et~al.}{2009}]{Dekel2009}
{Dekel} A.,  et~al., 2009, \mn@doi [\nat] {10.1038/nature07648}, \href
  {https://ui.adsabs.harvard.edu/abs/2009Natur.457..451D} {457, 451}

\bibitem[\protect\citeauthoryear{{Dolag}, {Borgani}, {Murante}  \&
  {Springel}}{{Dolag} et~al.}{2009}]{Dolag2009}
{Dolag} K.,  {Borgani} S.,  {Murante} G.,   {Springel} V.,  2009, \mn@doi
  [\mnras] {10.1111/j.1365-2966.2009.15034.x}, \href
  {https://ui.adsabs.harvard.edu/abs/2009MNRAS.399..497D} {399, 497}

\bibitem[\protect\citeauthoryear{{Donnari} et~al.,}{{Donnari}
  et~al.}{2019}]{Donnari2019}
{Donnari} M.,  et~al., 2019, \mn@doi [\mnras] {10.1093/mnras/stz712}, \href
  {https://ui.adsabs.harvard.edu/abs/2019MNRAS.485.4817D} {485, 4817}

\bibitem[\protect\citeauthoryear{{Du}, {Ho}, {Debattista}, {Pillepich},
  {Nelson}, {Hernquist}  \& {Weinberger}}{{Du} et~al.}{2021}]{Du2021}
{Du} M.,  {Ho} L.~C.,  {Debattista} V.~P.,  {Pillepich} A.,  {Nelson} D.,
  {Hernquist} L.,   {Weinberger} R.,  2021, arXiv e-prints, \href
  {https://ui.adsabs.harvard.edu/abs/2021arXiv210112373D} {p. arXiv:2101.12373}

\bibitem[\protect\citeauthoryear{{Dubois}, {Peirani}, {Pichon}, {Devriendt},
  {Gavazzi}, {Welker}  \& {Volonteri}}{{Dubois} et~al.}{2016}]{Dubois2016}
{Dubois} Y.,  {Peirani} S.,  {Pichon} C.,  {Devriendt} J.,  {Gavazzi} R.,
  {Welker} C.,   {Volonteri} M.,  2016, \mn@doi [\mnras]
  {10.1093/mnras/stw2265}, \href
  {https://ui.adsabs.harvard.edu/abs/2016MNRAS.463.3948D} {463, 3948}

\bibitem[\protect\citeauthoryear{{Emsellem} et~al.,}{{Emsellem}
  et~al.}{2011}]{Emsellem2011}
{Emsellem} E.,  et~al., 2011, \mn@doi [\mnras]
  {10.1111/j.1365-2966.2011.18496.x}, \href
  {https://ui.adsabs.harvard.edu/abs/2011MNRAS.414..888E} {414, 888}

\bibitem[\protect\citeauthoryear{{Engler} et~al.,}{{Engler}
  et~al.}{2021}]{Engler2020}
{Engler} C.,  et~al., 2021, \mn@doi [\mnras] {10.1093/mnras/staa3505}, \href
  {https://ui.adsabs.harvard.edu/abs/2021MNRAS.500.3957E} {500, 3957}

\bibitem[\protect\citeauthoryear{{Faber} et~al.,}{{Faber}
  et~al.}{2007}]{Faber2007}
{Faber} S.~M.,  et~al., 2007, \mn@doi [\apj] {10.1086/519294}, \href
  {https://ui.adsabs.harvard.edu/abs/2007ApJ...665..265F} {665, 265}

\bibitem[\protect\citeauthoryear{{Ferr{\'e}-Mateu}, {Mezcua}, {Trujillo},
  {Balcells}  \& {van den Bosch}}{{Ferr{\'e}-Mateu}
  et~al.}{2015}]{Ferre-Mateu2015}
{Ferr{\'e}-Mateu} A.,  {Mezcua} M.,  {Trujillo} I.,  {Balcells} M.,   {van den
  Bosch} R. C.~E.,  2015, \mn@doi [\apj] {10.1088/0004-637X/808/1/79}, \href
  {https://ui.adsabs.harvard.edu/abs/2015ApJ...808...79F} {808, 79}

\bibitem[\protect\citeauthoryear{{Ferr{\'e}-Mateu}, {Trujillo},
  {Mart{\'\i}n-Navarro}, {Vazdekis}, {Mezcua}, {Balcells}  \&
  {Dom{\'\i}nguez}}{{Ferr{\'e}-Mateu} et~al.}{2017}]{FerreMateu17}
{Ferr{\'e}-Mateu} A.,  {Trujillo} I.,  {Mart{\'\i}n-Navarro} I.,  {Vazdekis}
  A.,  {Mezcua} M.,  {Balcells} M.,   {Dom{\'\i}nguez} L.,  2017, \mn@doi
  [\mnras] {10.1093/mnras/stx171}, \href
  {https://ui.adsabs.harvard.edu/abs/2017MNRAS.467.1929F} {467, 1929}

\bibitem[\protect\citeauthoryear{{Ferr{\'e}-Mateu}, {Forbes}, {Romanowsky},
  {Janz}  \& {Dixon}}{{Ferr{\'e}-Mateu} et~al.}{2018}]{FerreMateu2018}
{Ferr{\'e}-Mateu} A.,  {Forbes} D.~A.,  {Romanowsky} A.~J.,  {Janz} J.,
  {Dixon} C.,  2018, \mn@doi [\mnras] {10.1093/mnras/stx2442}, \href
  {https://ui.adsabs.harvard.edu/abs/2018MNRAS.473.1819F} {473, 1819}

\bibitem[\protect\citeauthoryear{{Ferr{\'e}-Mateu}, {Durr{\'e}}, {Forbes},
  {Romanowsky}, {Alabi}, {Brodie}  \& {McDermid}}{{Ferr{\'e}-Mateu}
  et~al.}{2021}]{Ferre-Mateu2021}
{Ferr{\'e}-Mateu} A.,  {Durr{\'e}} M.,  {Forbes} D.~A.,  {Romanowsky} A.~J.,
  {Alabi} A.,  {Brodie} J.~P.,   {McDermid} R.~M.,  2021, \mn@doi [\mnras]
  {10.1093/mnras/stab805}, \href
  {https://ui.adsabs.harvard.edu/abs/2021MNRAS.503.5455F} {503, 5455}

\bibitem[\protect\citeauthoryear{{Furlong} et~al.,}{{Furlong}
  et~al.}{2017}]{Furlong2017}
{Furlong} M.,  et~al., 2017, \mn@doi [\mnras] {10.1093/mnras/stw2740}, \href
  {https://ui.adsabs.harvard.edu/abs/2017MNRAS.465..722F} {465, 722}

\bibitem[\protect\citeauthoryear{{Gallazzi}, {Charlot}, {Brinchmann}, {White}
  \& {Tremonti}}{{Gallazzi} et~al.}{2005}]{Gallazzi2005}
{Gallazzi} A.,  {Charlot} S.,  {Brinchmann} J.,  {White} S. D.~M.,   {Tremonti}
  C.~A.,  2005, \mn@doi [\mnras] {10.1111/j.1365-2966.2005.09321.x}, \href
  {https://ui.adsabs.harvard.edu/abs/2005MNRAS.362...41G} {362, 41}

\bibitem[\protect\citeauthoryear{{Genel} et~al.,}{{Genel}
  et~al.}{2018}]{Genel2018}
{Genel} S.,  et~al., 2018, \mn@doi [\mnras] {10.1093/mnras/stx3078}, \href
  {https://ui.adsabs.harvard.edu/abs/2018MNRAS.474.3976G} {474, 3976}

\bibitem[\protect\citeauthoryear{{Gonz{\'a}lez Delgado} et~al.,}{{Gonz{\'a}lez
  Delgado} et~al.}{2014}]{GonzalezDelgado2014}
{Gonz{\'a}lez Delgado} R.~M.,  et~al., 2014, \mn@doi [\apjl]
  {10.1088/2041-8205/791/1/L16}, \href
  {https://ui.adsabs.harvard.edu/abs/2014ApJ...791L..16G} {791, L16}

\bibitem[\protect\citeauthoryear{{Hilz}, {Naab}  \& {Ostriker}}{{Hilz}
  et~al.}{2013}]{Hilz2013}
{Hilz} M.,  {Naab} T.,   {Ostriker} J.~P.,  2013, \mn@doi [\mnras]
  {10.1093/mnras/sts501}, \href
  {https://ui.adsabs.harvard.edu/abs/2013MNRAS.429.2924H} {429, 2924}

\bibitem[\protect\citeauthoryear{{Hopkins}, {Hernquist}, {Cox}, {Keres}  \&
  {Wuyts}}{{Hopkins} et~al.}{2009}]{Hopkins2009}
{Hopkins} P.~F.,  {Hernquist} L.,  {Cox} T.~J.,  {Keres} D.,   {Wuyts} S.,
  2009, \mn@doi [\apj] {10.1088/0004-637X/691/2/1424}, \href
  {https://ui.adsabs.harvard.edu/abs/2009ApJ...691.1424H} {691, 1424}

\bibitem[\protect\citeauthoryear{{Hopkins}, {Wellons}, {Angles-Alcazar},
  {Faucher-Giguere}  \& {Grudic}}{{Hopkins} et~al.}{2021}]{Hopkins2021}
{Hopkins} P.~F.,  {Wellons} S.,  {Angles-Alcazar} D.,  {Faucher-Giguere} C.-A.,
    {Grudic} M.~Y.,  2021, arXiv e-prints, \href
  {https://ui.adsabs.harvard.edu/abs/2021arXiv210310444H} {p. arXiv:2103.10444}

\bibitem[\protect\citeauthoryear{{Kang} \& {Lee}}{{Kang} \&
  {Lee}}{2021}]{Kang2021}
{Kang} J.,  {Lee} M.~G.,  2021, arXiv e-prints, \href
  {https://ui.adsabs.harvard.edu/abs/2021arXiv210400672K} {p. arXiv:2104.00672}

\bibitem[\protect\citeauthoryear{{Lucatelli} \& {Ferrari}}{{Lucatelli} \&
  {Ferrari}}{2019}]{Lucatelli2019}
{Lucatelli} G.,  {Ferrari} F.,  2019, \mn@doi [\mnras] {10.1093/mnras/stz2154},
  \href {https://ui.adsabs.harvard.edu/abs/2019MNRAS.489.1161L} {489, 1161}

\bibitem[\protect\citeauthoryear{{Mart{\'\i}n-Navarro}, {La Barbera},
  {Vazdekis}, {Ferr{\'e}-Mateu}, {Trujillo}  \&
  {Beasley}}{{Mart{\'\i}n-Navarro} et~al.}{2015}]{martin-navarro15}
{Mart{\'\i}n-Navarro} I.,  {La Barbera} F.,  {Vazdekis} A.,  {Ferr{\'e}-Mateu}
  A.,  {Trujillo} I.,   {Beasley} M.~A.,  2015, \mn@doi [\mnras]
  {10.1093/mnras/stv1022}, \href
  {https://ui.adsabs.harvard.edu/abs/2015MNRAS.451.1081M} {451, 1081}

\bibitem[\protect\citeauthoryear{{Mart{\'\i}n-Navarro}, {Vazdekis},
  {Falc{\'o}n-Barroso}, {La Barbera}, {Y{\i}ld{\i}r{\i}m}  \& {van de
  Ven}}{{Mart{\'\i}n-Navarro} et~al.}{2018}]{Martin-Navarro2018a}
{Mart{\'\i}n-Navarro} I.,  {Vazdekis} A.,  {Falc{\'o}n-Barroso} J.,  {La
  Barbera} F.,  {Y{\i}ld{\i}r{\i}m} A.,   {van de Ven} G.,  2018, \mn@doi
  [\mnras] {10.1093/mnras/stx3346}, \href
  {https://ui.adsabs.harvard.edu/abs/2018MNRAS.475.3700M} {475, 3700}

\bibitem[\protect\citeauthoryear{{Mart{\'\i}n-Navarro}, {van de Ven}  \&
  {Y{\i}ld{\i}r{\i}m}}{{Mart{\'\i}n-Navarro} et~al.}{2019}]{Martin-Navarro2019}
{Mart{\'\i}n-Navarro} I.,  {van de Ven} G.,   {Y{\i}ld{\i}r{\i}m} A.,  2019,
  \mn@doi [\mnras] {10.1093/mnras/stz1544}, \href
  {https://ui.adsabs.harvard.edu/abs/2019MNRAS.487.4939M} {487, 4939}

\bibitem[\protect\citeauthoryear{{Naab}, {Johansson}  \& {Ostriker}}{{Naab}
  et~al.}{2009}]{Naab09}
{Naab} T.,  {Johansson} P.~H.,   {Ostriker} J.~P.,  2009, \mn@doi [\apjl]
  {10.1088/0004-637X/699/2/L178}, \href
  {https://ui.adsabs.harvard.edu/abs/2009ApJ...699L.178N} {699, L178}

\bibitem[\protect\citeauthoryear{{Naiman} et~al.,}{{Naiman}
  et~al.}{2018}]{Naiman2018}
{Naiman} J.~P.,  et~al., 2018, \mn@doi [\mnras] {10.1093/mnras/sty618}, \href
  {https://ui.adsabs.harvard.edu/abs/2018MNRAS.477.1206N} {477, 1206}

\bibitem[\protect\citeauthoryear{{Nelson} et~al.,}{{Nelson}
  et~al.}{2018}]{nelson18}
{Nelson} D.,  et~al., 2018, \mn@doi [\mnras] {10.1093/mnras/stx3040}, \href
  {https://ui.adsabs.harvard.edu/abs/2018MNRAS.475..624N} {475, 624}

\bibitem[\protect\citeauthoryear{{Nelson} et~al.,}{{Nelson}
  et~al.}{2019a}]{Nelson2019b}
{Nelson} D.,  et~al., 2019a, \mn@doi [Computational Astrophysics and Cosmology]
  {10.1186/s40668-019-0028-x}, \href
  {https://ui.adsabs.harvard.edu/abs/2019ComAC...6....2N} {6, 2}

\bibitem[\protect\citeauthoryear{{Nelson} et~al.,}{{Nelson}
  et~al.}{2019b}]{Nelson2019a}
{Nelson} D.,  et~al., 2019b, \mn@doi [\mnras] {10.1093/mnras/stz2306}, \href
  {https://ui.adsabs.harvard.edu/abs/2019MNRAS.490.3234N} {490, 3234}

\bibitem[\protect\citeauthoryear{{Newman}, {Ellis}, {Bundy}  \&
  {Treu}}{{Newman} et~al.}{2012}]{Newman2012}
{Newman} A.~B.,  {Ellis} R.~S.,  {Bundy} K.,   {Treu} T.,  2012, \mn@doi [\apj]
  {10.1088/0004-637X/746/2/162}, \href
  {https://ui.adsabs.harvard.edu/abs/2012ApJ...746..162N} {746, 162}

\bibitem[\protect\citeauthoryear{{Oser}, {Ostriker}, {Naab}, {Johansson}  \&
  {Burkert}}{{Oser} et~al.}{2010}]{Oser2010}
{Oser} L.,  {Ostriker} J.~P.,  {Naab} T.,  {Johansson} P.~H.,   {Burkert} A.,
  2010, \mn@doi [\apj] {10.1088/0004-637X/725/2/2312}, \href
  {https://ui.adsabs.harvard.edu/abs/2010ApJ...725.2312O} {725, 2312}

\bibitem[\protect\citeauthoryear{{Oser}, {Naab}, {Ostriker}  \&
  {Johansson}}{{Oser} et~al.}{2012}]{Oser2012}
{Oser} L.,  {Naab} T.,  {Ostriker} J.~P.,   {Johansson} P.~H.,  2012, \mn@doi
  [\apj] {10.1088/0004-637X/744/1/63}, \href
  {https://ui.adsabs.harvard.edu/abs/2012ApJ...744...63O} {744, 63}

\bibitem[\protect\citeauthoryear{{Panter}, {Jimenez}, {Heavens}  \&
  {Charlot}}{{Panter} et~al.}{2008}]{Panter2008}
{Panter} B.,  {Jimenez} R.,  {Heavens} A.~F.,   {Charlot} S.,  2008, \mn@doi
  [\mnras] {10.1111/j.1365-2966.2008.13981.x}, \href
  {https://ui.adsabs.harvard.edu/abs/2008MNRAS.391.1117P} {391, 1117}

\bibitem[\protect\citeauthoryear{{Peralta de Arriba}, {Quilis}, {Trujillo},
  {Cebri{\'a}n}  \& {Balcells}}{{Peralta de Arriba} et~al.}{2016}]{Peralta2016}
{Peralta de Arriba} L.,  {Quilis} V.,  {Trujillo} I.,  {Cebri{\'a}n} M.,
  {Balcells} M.,  2016, \mn@doi [\mnras] {10.1093/mnras/stw1240}, \href
  {https://ui.adsabs.harvard.edu/abs/2016MNRAS.461..156P} {461, 156}

\bibitem[\protect\citeauthoryear{{Pillepich} et~al.,}{{Pillepich}
  et~al.}{2018a}]{Pillepich2018a}
{Pillepich} A.,  et~al., 2018a, \mn@doi [\mnras] {10.1093/mnras/stx2656}, \href
  {https://ui.adsabs.harvard.edu/abs/2018MNRAS.473.4077P} {473, 4077}

\bibitem[\protect\citeauthoryear{{Pillepich} et~al.,}{{Pillepich}
  et~al.}{2018b}]{Pillepich18b}
{Pillepich} A.,  et~al., 2018b, \mn@doi [\mnras] {10.1093/mnras/stx3112}, \href
  {https://ui.adsabs.harvard.edu/abs/2018MNRAS.475..648P} {475, 648}

\bibitem[\protect\citeauthoryear{{Pillepich} et~al.,}{{Pillepich}
  et~al.}{2019a}]{Pillepich19}
{Pillepich} A.,  et~al., 2019a, \mn@doi [\mnras] {10.1093/mnras/stz2338}, \href
  {https://ui.adsabs.harvard.edu/abs/2019MNRAS.490.3196P} {490, 3196}

\bibitem[\protect\citeauthoryear{{Pillepich} et~al.,}{{Pillepich}
  et~al.}{2019b}]{Pillepich2019}
{Pillepich} A.,  et~al., 2019b, \mn@doi [\mnras] {10.1093/mnras/stz2338}, \href
  {https://ui.adsabs.harvard.edu/abs/2019MNRAS.490.3196P} {490, 3196}

\bibitem[\protect\citeauthoryear{{Planck Collaboration} et~al.,}{{Planck
  Collaboration} et~al.}{2016}]{Planck2016}
{Planck Collaboration} et~al., 2016, \mn@doi [\aap]
  {10.1051/0004-6361/201525830}, \href
  {https://ui.adsabs.harvard.edu/abs/2016A&A...594A..13P} {594, A13}

\bibitem[\protect\citeauthoryear{{Poggianti} et~al.,}{{Poggianti}
  et~al.}{2013a}]{Poggianti2013a}
{Poggianti} B.~M.,  et~al., 2013a, \mn@doi [\apj] {10.1088/0004-637X/762/2/77},
  \href {https://ui.adsabs.harvard.edu/abs/2013ApJ...762...77P} {762, 77}

\bibitem[\protect\citeauthoryear{{Poggianti}, {Moretti}, {Calvi}, {D'Onofrio},
  {Valentinuzzi}, {Fritz}  \& {Renzini}}{{Poggianti}
  et~al.}{2013b}]{Poggianti2013b}
{Poggianti} B.~M.,  {Moretti} A.,  {Calvi} R.,  {D'Onofrio} M.,  {Valentinuzzi}
  T.,  {Fritz} J.,   {Renzini} A.,  2013b, \mn@doi [\apj]
  {10.1088/0004-637X/777/2/125}, \href
  {https://ui.adsabs.harvard.edu/abs/2013ApJ...777..125P} {777, 125}

\bibitem[\protect\citeauthoryear{{Popping} et~al.,}{{Popping}
  et~al.}{2019}]{Popping2019}
{Popping} G.,  et~al., 2019, \mn@doi [\apj] {10.3847/1538-4357/ab30f2}, \href
  {https://ui.adsabs.harvard.edu/abs/2019ApJ...882..137P} {882, 137}

\bibitem[\protect\citeauthoryear{{Pulsoni}, {Gerhard}, {Arnaboldi},
  {Pillepich}, {Nelson}, {Hernquist}  \& {Springel}}{{Pulsoni}
  et~al.}{2020}]{Pulsoni2020}
{Pulsoni} C.,  {Gerhard} O.,  {Arnaboldi} M.,  {Pillepich} A.,  {Nelson} D.,
  {Hernquist} L.,   {Springel} V.,  2020, \mn@doi [\aap]
  {10.1051/0004-6361/202038253}, \href
  {https://ui.adsabs.harvard.edu/abs/2020A&A...641A..60P} {641, A60}

\bibitem[\protect\citeauthoryear{{Quilis} \& {Trujillo}}{{Quilis} \&
  {Trujillo}}{2013}]{Quilis2013}
{Quilis} V.,  {Trujillo} I.,  2013, \mn@doi [\apjl]
  {10.1088/2041-8205/773/1/L8}, \href
  {https://ui.adsabs.harvard.edu/abs/2013ApJ...773L...8Q} {773, L8}

\bibitem[\protect\citeauthoryear{{Rodriguez-Gomez} et~al.,}{{Rodriguez-Gomez}
  et~al.}{2015}]{Rodriguez-Gomez2015}
{Rodriguez-Gomez} V.,  et~al., 2015, \mn@doi [\mnras] {10.1093/mnras/stv264},
  \href {https://ui.adsabs.harvard.edu/abs/2015MNRAS.449...49R} {449, 49}

\bibitem[\protect\citeauthoryear{{Rodriguez-Gomez} et~al.,}{{Rodriguez-Gomez}
  et~al.}{2016}]{rodriguez-gomez2016}
{Rodriguez-Gomez} V.,  et~al., 2016, \mn@doi [\mnras] {10.1093/mnras/stw456},
  \href {https://ui.adsabs.harvard.edu/abs/2016MNRAS.458.2371R} {458, 2371}

\bibitem[\protect\citeauthoryear{{Saulder}, {van den Bosch}  \&
  {Mieske}}{{Saulder} et~al.}{2015}]{Saulder2015}
{Saulder} C.,  {van den Bosch} R. C.~E.,   {Mieske} S.,  2015, \mn@doi [\aap]
  {10.1051/0004-6361/201425472}, \href
  {https://ui.adsabs.harvard.edu/abs/2015A&A...578A.134S} {578, A134}

\bibitem[\protect\citeauthoryear{{Schaye} et~al.,}{{Schaye}
  et~al.}{2015}]{Schaye2015}
{Schaye} J.,  et~al., 2015, \mn@doi [\mnras] {10.1093/mnras/stu2058}, \href
  {https://ui.adsabs.harvard.edu/abs/2015MNRAS.446..521S} {446, 521}

\bibitem[\protect\citeauthoryear{{Schreiber} et~al.,}{{Schreiber}
  et~al.}{2018}]{Schreiber2018}
{Schreiber} C.,  et~al., 2018, \mn@doi [\aap] {10.1051/0004-6361/201833070},
  \href {https://ui.adsabs.harvard.edu/abs/2018A&A...618A..85S} {618, A85}

\bibitem[\protect\citeauthoryear{{Spiniello} et~al.,}{{Spiniello}
  et~al.}{2021}]{spiniello2021}
{Spiniello} C.,  et~al., 2021, \mn@doi [\aap] {10.1051/0004-6361/202038936},
  \href {https://ui.adsabs.harvard.edu/abs/2021A&A...646A..28S} {646, A28}

\bibitem[\protect\citeauthoryear{{Spolaor}, {Kobayashi}, {Forbes}, {Couch}  \&
  {Hau}}{{Spolaor} et~al.}{2010}]{Spoalor2010}
{Spolaor} M.,  {Kobayashi} C.,  {Forbes} D.~A.,  {Couch} W.~J.,   {Hau} G.
  K.~T.,  2010, \mn@doi [\mnras] {10.1111/j.1365-2966.2010.17080.x}, \href
  {https://ui.adsabs.harvard.edu/abs/2010MNRAS.408..272S} {408, 272}

\bibitem[\protect\citeauthoryear{{Springel}, {White}, {Tormen}  \&
  {Kauffmann}}{{Springel} et~al.}{2001}]{Springel2001}
{Springel} V.,  {White} S. D.~M.,  {Tormen} G.,   {Kauffmann} G.,  2001,
  \mn@doi [\mnras] {10.1046/j.1365-8711.2001.04912.x}, \href
  {https://ui.adsabs.harvard.edu/abs/2001MNRAS.328..726S} {328, 726}

\bibitem[\protect\citeauthoryear{{Springel} et~al.,}{{Springel}
  et~al.}{2018}]{Springel18}
{Springel} V.,  et~al., 2018, \mn@doi [\mnras] {10.1093/mnras/stx3304}, \href
  {https://ui.adsabs.harvard.edu/abs/2018MNRAS.475..676S} {475, 676}

\bibitem[\protect\citeauthoryear{{Stringer}, {Trujillo}, {Dalla Vecchia}  \&
  {Martinez-Valpuesta}}{{Stringer} et~al.}{2015}]{Stringer2015}
{Stringer} M.,  {Trujillo} I.,  {Dalla Vecchia} C.,   {Martinez-Valpuesta} I.,
  2015, \mn@doi [\mnras] {10.1093/mnras/stv455}, \href
  {https://ui.adsabs.harvard.edu/abs/2015MNRAS.449.2396S} {449, 2396}

\bibitem[\protect\citeauthoryear{{Szomoru}, {Franx}  \& {van Dokkum}}{{Szomoru}
  et~al.}{2012}]{Sozomoru2012}
{Szomoru} D.,  {Franx} M.,   {van Dokkum} P.~G.,  2012, \mn@doi [\apj]
  {10.1088/0004-637X/749/2/121}, \href
  {https://ui.adsabs.harvard.edu/abs/2012ApJ...749..121S} {749, 121}

\bibitem[\protect\citeauthoryear{{Taylor}, {Franx}, {Glazebrook}, {Brinchmann},
  {van der Wel}  \& {van Dokkum}}{{Taylor} et~al.}{2010}]{Taylor2010}
{Taylor} E.~N.,  {Franx} M.,  {Glazebrook} K.,  {Brinchmann} J.,  {van der Wel}
  A.,   {van Dokkum} P.~G.,  2010, \mn@doi [\apj]
  {10.1088/0004-637X/720/1/723}, \href
  {https://ui.adsabs.harvard.edu/abs/2010ApJ...720..723T} {720, 723}

\bibitem[\protect\citeauthoryear{{Tortora} et~al.,}{{Tortora}
  et~al.}{2016}]{Tortora2016}
{Tortora} C.,  et~al., 2016, \mn@doi [\mnras] {10.1093/mnras/stw184}, \href
  {https://ui.adsabs.harvard.edu/abs/2016MNRAS.457.2845T} {457, 2845}

\bibitem[\protect\citeauthoryear{Trayford et~al.,}{Trayford
  et~al.}{2017}]{Trayford2017OpticalSKIRT}
Trayford J.~W.,  et~al., 2017, \mn@doi [Monthly Notices of the Royal
  Astronomical Society] {10.1093/mnras/stx1051}, 470, 771

\bibitem[\protect\citeauthoryear{{Trujillo}, {Conselice}, {Bundy}, {Cooper},
  {Eisenhardt}  \& {Ellis}}{{Trujillo} et~al.}{2007}]{Trujillo2007}
{Trujillo} I.,  {Conselice} C.~J.,  {Bundy} K.,  {Cooper} M.~C.,  {Eisenhardt}
  P.,   {Ellis} R.~S.,  2007, \mn@doi [\mnras]
  {10.1111/j.1365-2966.2007.12388.x}, \href
  {https://ui.adsabs.harvard.edu/abs/2007MNRAS.382..109T} {382, 109}

\bibitem[\protect\citeauthoryear{{Trujillo}, {Cenarro}, {de
  Lorenzo-C{\'a}ceres}, {Vazdekis}, {de la Rosa}  \& {Cava}}{{Trujillo}
  et~al.}{2009}]{Trujillo2009}
{Trujillo} I.,  {Cenarro} A.~J.,  {de Lorenzo-C{\'a}ceres} A.,  {Vazdekis} A.,
  {de la Rosa} I.~G.,   {Cava} A.,  2009, \mn@doi [\apjl]
  {10.1088/0004-637X/692/2/L118}, \href
  {https://ui.adsabs.harvard.edu/abs/2009ApJ...692L.118T} {692, L118}

\bibitem[\protect\citeauthoryear{{Trujillo}, {Ferr{\'e}-Mateu}, {Balcells},
  {Vazdekis}  \& {S{\'a}nchez-Bl{\'a}zquez}}{{Trujillo}
  et~al.}{2014}]{Trujillo2014}
{Trujillo} I.,  {Ferr{\'e}-Mateu} A.,  {Balcells} M.,  {Vazdekis} A.,
  {S{\'a}nchez-Bl{\'a}zquez} P.,  2014, \mn@doi [\apjl]
  {10.1088/2041-8205/780/2/L20}, \href
  {https://ui.adsabs.harvard.edu/abs/2014ApJ...780L..20T} {780, L20}

\bibitem[\protect\citeauthoryear{{Valentino} et~al.,}{{Valentino}
  et~al.}{2020}]{Valentino2020}
{Valentino} F.,  et~al., 2020, \mn@doi [\apj] {10.3847/1538-4357/ab64dc}, \href
  {https://ui.adsabs.harvard.edu/abs/2020ApJ...889...93V} {889, 93}

\bibitem[\protect\citeauthoryear{{Vogelsberger}, {Marinacci}, {Torrey}  \&
  {Puchwein}}{{Vogelsberger} et~al.}{2020}]{Vogelsberger2020}
{Vogelsberger} M.,  {Marinacci} F.,  {Torrey} P.,   {Puchwein} E.,  2020,
  \mn@doi [Nature Reviews Physics] {10.1038/s42254-019-0127-2}, \href
  {https://ui.adsabs.harvard.edu/abs/2020NatRP...2...42V} {2, 42}

\bibitem[\protect\citeauthoryear{{Weinberger} et~al.,}{{Weinberger}
  et~al.}{2017}]{Weinberger2017}
{Weinberger} R.,  et~al., 2017, \mn@doi [\mnras] {10.1093/mnras/stw2944}, \href
  {https://ui.adsabs.harvard.edu/abs/2017MNRAS.465.3291W} {465, 3291}

\bibitem[\protect\citeauthoryear{{Weinberger}, {Springel}  \&
  {Pakmor}}{{Weinberger} et~al.}{2020}]{Weinberger2020}
{Weinberger} R.,  {Springel} V.,   {Pakmor} R.,  2020, \mn@doi [\apjs]
  {10.3847/1538-4365/ab908c}, \href
  {https://ui.adsabs.harvard.edu/abs/2020ApJS..248...32W} {248, 32}

\bibitem[\protect\citeauthoryear{{Wellons} et~al.,}{{Wellons}
  et~al.}{2015}]{wellons15}
{Wellons} S.,  et~al., 2015, \mn@doi [\mnras] {10.1093/mnras/stv303}, \href
  {https://ui.adsabs.harvard.edu/abs/2015MNRAS.449..361W} {449, 361}

\bibitem[\protect\citeauthoryear{{Wellons} et~al.,}{{Wellons}
  et~al.}{2016}]{wellons16}
{Wellons} S.,  et~al., 2016, \mn@doi [\mnras] {10.1093/mnras/stv2738}, \href
  {https://ui.adsabs.harvard.edu/abs/2016MNRAS.456.1030W} {456, 1030}

\bibitem[\protect\citeauthoryear{{Werner}, {Lakhchaura}, {Canning}, {Gaspari}
  \& {Simionescu}}{{Werner} et~al.}{2018}]{Werner2018}
{Werner} N.,  {Lakhchaura} K.,  {Canning} R.~E.~A.,  {Gaspari} M.,
  {Simionescu} A.,  2018, \mn@doi [\mnras] {10.1093/mnras/sty862}, \href
  {https://ui.adsabs.harvard.edu/abs/2018MNRAS.477.3886W} {477, 3886}

\bibitem[\protect\citeauthoryear{{Wetzel}, {Tinker}  \& {Conroy}}{{Wetzel}
  et~al.}{2012}]{Wetzel2012}
{Wetzel} A.~R.,  {Tinker} J.~L.,   {Conroy} C.,  2012, \mn@doi [\mnras]
  {10.1111/j.1365-2966.2012.21188.x}, \href
  {https://ui.adsabs.harvard.edu/abs/2012MNRAS.424..232W} {424, 232}

\bibitem[\protect\citeauthoryear{{Y{\i}ld{\i}r{\i}m}, {van den Bosch}, {van de
  Ven}, {Husemann}, {Lyubenova}, {Walsh}, {Gebhardt}  \&
  {G{\"u}ltekin}}{{Y{\i}ld{\i}r{\i}m} et~al.}{2015}]{Yildirim2015}
{Y{\i}ld{\i}r{\i}m} A.,  {van den Bosch} R. C.~E.,  {van de Ven} G.,
  {Husemann} B.,  {Lyubenova} M.,  {Walsh} J.~L.,  {Gebhardt} K.,
  {G{\"u}ltekin} K.,  2015, \mn@doi [\mnras] {10.1093/mnras/stv1381}, \href
  {https://ui.adsabs.harvard.edu/abs/2015MNRAS.452.1792Y} {452, 1792}

\bibitem[\protect\citeauthoryear{{Y{\i}ld{\i}r{\i}m}, {van den Bosch}, {van de
  Ven}, {Mart{\'\i}n-Navarro}, {Walsh}, {Husemann}, {G{\"u}ltekin}  \&
  {Gebhardt}}{{Y{\i}ld{\i}r{\i}m} et~al.}{2017}]{yildirim17}
{Y{\i}ld{\i}r{\i}m} A.,  {van den Bosch} R. C.~E.,  {van de Ven} G.,
  {Mart{\'\i}n-Navarro} I.,  {Walsh} J.~L.,  {Husemann} B.,  {G{\"u}ltekin} K.,
    {Gebhardt} K.,  2017, \mn@doi [\mnras] {10.1093/mnras/stx732}, \href
  {https://ui.adsabs.harvard.edu/abs/2017MNRAS.468.4216Y} {468, 4216}

\bibitem[\protect\citeauthoryear{{Zanisi} et~al.,}{{Zanisi}
  et~al.}{2021}]{Zanisi2021}
{Zanisi} L.,  et~al., 2021, \mn@doi [\mnras] {10.1093/mnras/staa3864}, \href
  {https://ui.adsabs.harvard.edu/abs/2021MNRAS.501.4359Z} {501, 4359}

\bibitem[\protect\citeauthoryear{{Zibetti}, {Gallazzi}, {Hirschmann},
  {Consolandi}, {Falc{\'o}n-Barroso}, {van de Ven}  \& {Lyubenova}}{{Zibetti}
  et~al.}{2020}]{Zibetti2020}
{Zibetti} S.,  {Gallazzi} A.~R.,  {Hirschmann} M.,  {Consolandi} G.,
  {Falc{\'o}n-Barroso} J.,  {van de Ven} G.,   {Lyubenova} M.,  2020, \mn@doi
  [\mnras] {10.1093/mnras/stz3205}, \href
  {https://ui.adsabs.harvard.edu/abs/2020MNRAS.491.3562Z} {491, 3562}

\bibitem[\protect\citeauthoryear{{Zolotov} et~al.,}{{Zolotov}
  et~al.}{2015}]{Zolotov2015}
{Zolotov} A.,  et~al., 2015, \mn@doi [\mnras] {10.1093/mnras/stv740}, \href
  {https://ui.adsabs.harvard.edu/abs/2015MNRAS.450.2327Z} {450, 2327}

\bibitem[\protect\citeauthoryear{{van Dokkum} et~al.,}{{van Dokkum}
  et~al.}{2010}]{vanDokkum10}
{van Dokkum} P.~G.,  et~al., 2010, \mn@doi [\apj]
  {10.1088/0004-637X/709/2/1018}, \href
  {https://ui.adsabs.harvard.edu/abs/2010ApJ...709.1018V} {709, 1018}

\bibitem[\protect\citeauthoryear{{van den Bosch}, {Gebhardt}, {G{\"u}ltekin},
  {van de Ven}, {van der Wel}  \& {Walsh}}{{van den Bosch}
  et~al.}{2012}]{vandenBosch2012}
{van den Bosch} R. C.~E.,  {Gebhardt} K.,  {G{\"u}ltekin} K.,  {van de Ven} G.,
   {van der Wel} A.,   {Walsh} J.~L.,  2012, \mn@doi [\nat]
  {10.1038/nature11592}, \href
  {https://ui.adsabs.harvard.edu/abs/2012Natur.491..729V} {491, 729}

\bibitem[\protect\citeauthoryear{{van der Wel} et~al.,}{{van der Wel}
  et~al.}{2011}]{vanderWel2011}
{van der Wel} A.,  et~al., 2011, \mn@doi [\apj] {10.1088/0004-637X/730/1/38},
  \href {https://ui.adsabs.harvard.edu/abs/2011ApJ...730...38V} {730, 38}

\bibitem[\protect\citeauthoryear{{van der Wel} et~al.,}{{van der Wel}
  et~al.}{2014}]{vanderWel14}
{van der Wel} A.,  et~al., 2014, \mn@doi [\apj] {10.1088/0004-637X/788/1/28},
  \href {https://ui.adsabs.harvard.edu/abs/2014ApJ...788...28V} {788, 28}

\makeatother
\end{thebibliography}



\appendix

\section{Stellar populations ages}\label{app:stell_pop}
In this appendix, we show the stellar mass fraction contribution of stellar populations with different ages in the simulated galaxies.

\begin{figure}
    \centering
    \includegraphics[width=0.45\textwidth]{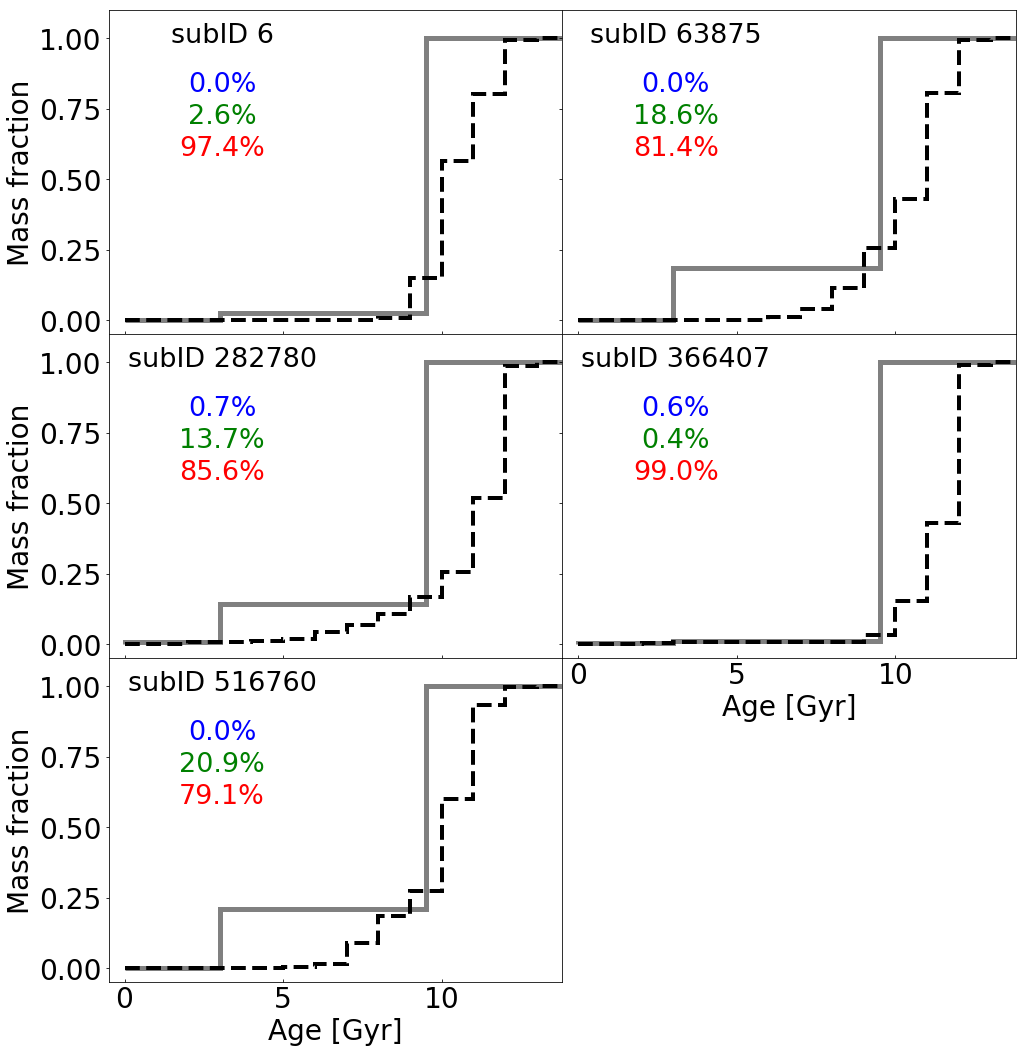}
    \caption{Cumulative stellar mass fraction as a function of age in stellar populations for each relic analogue candidate. Different lines represent cumulative histograms with different bin widths. Dashed black lines represent histograms with regular bin widths of 1 Gyr, while solid grey lines represent histograms with bin edges at particular age values: 0 Gyr, 3 Gyr, 9.5 Gyr an 13.6 Gyr ($z=0$, $z \sim 0.25$, $z \sim 1.5$ and $z \sim 20$). Coloured annotations show mass fraction contributions of young (blue, age $\leq$ 3 Gyr), intermediate (green, 3 Gyr $<$ age $\leq$ 9.5 Gyr) and old (red, 9.5 Gyr $<$ age) stellar populations.}
    \label{fig:stelpop}
\end{figure}


\bsp	
\label{lastpage}
\end{document}